\documentclass[twocolumn,showpacs,prb]{revtex4}
\usepackage{amsbsy,amssymb,amsmath,bm}
\usepackage{graphicx,color}
\usepackage{dcolumn}

\begin{document}

\title{Coulomb effects and hopping transport in
granular metals}
\author{I.~S.~Beloborodov,  A.~V.~Lopatin, and V.~M.~Vinokur}
\address{Materials Science Division, Argonne National
Laboratory, Argonne, Illinois 60439, USA}

\date{\today}

\begin{abstract}
We investigate effects of Coulomb interaction and hopping
transport in the insulator phase of granular metals and quantum
dot arrays.  We consider a spatially periodic as well as an
irregular array, including disorder in a form of a random on-site
electrostatic potential.  We study the Mott transition between the
insulating and metallic states in the regular system and find the
dependence of the Mott gap upon the intergranular coupling. The
conductivity of a strictly periodic array has an activation form
with the Mott gap as an activation energy. Considering irregular
systems we concentrate on the transport properties in the
dielectric, low coupling limit and derive the Efros-Shklovskii law
for hopping conductivity. In the irregular arrays electrostatic
disorder results in the finite density of states on the Fermi
level giving rise to the variable range hopping mechanism. We
develop a theory of tunneling through a chain of grains and
discuss in detail both elastic and inelastic cotunneling
mechanisms; the former dominates at very low temperatures and/or
very low applied electric fields, while the inelastic mechanism
controls tunneling at high temperature/fields. Our results are
obtained within the framework of the new technique based on the
mapping of quantum electronic problem onto the classical gas of
Coulomb charges. The processes of quantum tunnelling of real
electrons are represented in this technique as trajectories (world
lines) of charged classical particles in $d+1$ dimensions.  The
Mott gap is related to the dielectric susceptibility of the
Coulomb gas in the direction of the imaginary time axis.
\end{abstract}

\maketitle

\section{Introduction}

Electron transport in granular metals is a subject of intense
theoretical and experimental
research.\cite{experiment,Hendrich,Efetov02,Shklovskii04,BLV03,Universal,Feigelman04,kozub,Loh}.
The  interest is motivated  by the fact that granular metals
represent a unique tunable system that captures all the essential
features of general disordered systems and allows for the study
the effects of interference of strong Coulomb interaction and
disorder.   In particular, depending on the intergranular
conductance, granular metals exhibit the wide spectrum of
transport behaviors ranging from the hopping conductivity to those
typical to disordered Fermi liquids.

At present the metallic regime corresponding to strong
intergranular coupling where the Coulomb repulsion is
substantially screened is understood fairly
well\cite{Efetov02,BLV03}.  The key energy scale in this regime is
the inverse escape rate from the grain $\Gamma=g\delta$, where
$\delta$ is the mean level spacing in the individual granule, and
$g$ is the intergranular conductance. Generally, on the energy
scales larger than $\Gamma$ the granularity is essential, while at
lower energy scales the system resembles an ordinary disordered
metal with an effective diffusion coefficient determined by the
coupling between the grains. As a result at temperatures larger
than $\Gamma$  the conductivity exhibits the logarithmic
temperature dependence in all dimensions, while the low
temperature phase at $T <\Gamma$,~\cite{Efetov02} can be viewed as
the disordered Fermi liquid~\cite{BLV03,Universal} with
conductivity described by the
  Altshuler-Aronov-  \cite{Altshuler-Aronov}
 and localization type corrections.~\cite{GLK}

The dielectric regime where the conductivity is known to depend on
the temperature as~\cite{Abeles}:
  \begin{equation}
    \label{hopping} \sigma = \sigma_0\exp{[-(T_{\circ}/T)^{1/2}]},
  \end{equation}
with $\sigma_0$ being the high temperature conductivity and
$T_{\circ}$ being the characteristic energy scale defined below,
has been long posing a puzzle.  This form of the observed behavior
suggested the primary role of Coulomb correlations and several
models were proposed, yet the problem of conduction in granular
metals remained unsolved, see Ref.~[\onlinecite{pollak}] for a
review.  Building on the constructive critique of
Ref.~[\onlinecite{pollak}], the important development towards
understanding the observed behavior (\ref{hopping}) in granular
metals was made in Ref.~[\onlinecite{Shklovskii04}], where the
role of electrostatic disorder in formation of random potentials
on individual granules was fully recognized and explored.  This
suggested that the dependence (\ref{hopping}) can indeed be viewed
as Mott-Efros-Shklovskii variable range hopping in the presence of
the Coulomb gap~\cite{Efros}. The remaining problem however was
that in order to realize variable range hopping conductivity an
electron have to tunnel through the sequence of granules in one
hop and no mechanism that could provide such a tunneling process
was described.

In the Letter~[\onlinecite{BLV05}] we have reported the mechanism,
the multiple coherent co-tunneling, via which the correlated
hopping in a Coulomb gap occurs in granular conductors.  We
developed the technique allowing rigorous consideration of Coulomb
effects putting thus the theory of hopping transport in granular
conductors on a firm quantitative theoretical basis. In the
present work we investigate in detail the effects of the Coulomb
interaction on transport properties of granular metallic arrays
concentrating on the regime of low and intermediate intergranular
coupling.  We discuss both the model of a periodic granular array
as well as the irregular model that includes the random on-site
electrostatic potential.  We begin with the periodic model and
show that the regular array of metallic granules is a Mott
insulator at zero temperature, provided the tunneling conductance
is less than the critical value $g_c=(1/ \pi z) \ln(
E_0^c/\delta)$ where $z$ is the coordination number ( number of
neighbors for a site on the array ), $\delta$ is the mean energy
level spacing  and $ E_0^c$ is the Coulomb charging energy in a
single grain.  The Mott gap, the characteristic property of the
insulating state, decreases with the increase of the intergranular
conductance and disappears when $g$ riches the critical value
$g_c$.

We develop an approach to the insulating state of granular
conductors based on the mapping the original quantum model
describing electrons in the granular system on the classical gas
of the interacting Coulomb charges. More specifically, we first
absorb the Coulomb interaction into the phase field, similarly to
Ref.~[\onlinecite{AES}] and next map the quantum problem that
involves functional integration over the phase fields onto the
classical model that has an extra time dimension -- in an analogy
with the method of Ref.~[\onlinecite{Schmid}].  The classical
charges of the emerging effective model are connected via the
electron world lines representing graphically tunneling processes
and form necklace-like  loops. In the insulating state only short
loops are present; this means that electrons and holes form bound
states and only virtual tunnellings over the short distances are
allowed. In the metallic phase the loops of the infinite size are
present meaning  that electrons and holes are not bound together.
We show that the Mott gap in the insulating state is given by the
time-axis component of the inverse dielectric susceptibility of
the classical charged gas.

Mott transition in a periodic granular array resembles the Mott
transition in the Hubbard model~\cite{Kotliar}.  However, there is
an important difference between the two. Namely, in granular
metals, even in the case of the periodic granular samples, the
electron motion inside the grains as well as on the scales
exceeding the intergranular distance is diffusive, while the
Hubbard model assumes that electrons move through a periodic
lattice that allows to label their states by quasimomenta. The
model of a granular metal also has an extra physical parameter
$\delta,$ the mean energy levels spacing in a single grain, that
has no analog in the Hubbard model.  Yet, the physics of the
transition is similar in both cases, and one may expect that the
methods developed for the study of the Mott transition in the
system of strongly interacting electrons on the lattice (such as
dynamical mean field theory~\cite{Kotliar}, for example) can be
applied to the granular metals as well.  At present, however, the
description of the Mott gap in granular arrays in the very
vicinity of the Mott transition is not available, in particular
the order of the transition remains an unresolved problem.

The spatially regular granular arrays, including technologically
important nanocrystals of quantum dots, are now the subject of the
extensive experimental study. One might expect that periodic
arrays would show activation transport, characteristic for the
systems with the hard gap in the electron spectrum.  However, even
seemingly perfectly periodic arrays
\cite{Hendrich,Yakimov,Hendrich1} do not follow the expected
activation conductivity behavior. Instead, the experimentally
observed conductivity follows the Efros-Shklovskii (ES) law
expressed by Eq. (\ref{hopping}).  This behavior indicates that
electrostatic disorder plays a primary role in these systems.

Dependence (\ref{hopping}) cannot be observed in the strictly
periodic structures without disorder\cite{footenote1}, where the
hard gap in the excitation spectrum results in the activation
behavior of the conductivity.  The two necessary ingredients for
the variable-range-hopping type(VRH) conductivity are: (i) the
finite density of states at energies close to Fermi level and (ii)
the exponentially falling probability of tunneling between the
states close to the Fermi level. Furthermore, the transformation
of the usual Mott's  VRH dependence into the Efros-Shklovskii law
reflects the presence of the so-called soft gap in the electron
spectrum resulting from the strong long range Coulomb
correlations. The problem of hopping transport in granular
conductors is thus two-fold:  (i) to explain the origin of the
finite density of states near the Fermi-level and the role of
Coulomb correlations and (ii) to elucidate the mechanism of
tunneling through the dense array of metallic grains.

Earlier attempts to explain the behavior (\ref{hopping}) were
based on the assumption of the large dispersion in the granule's
size.~\cite{Abeles} Moreover, one more strong assumption was
necessary in order to arrive at formula (\ref{hopping}) that the
distance between the grains and the grain sizes are not
statistically independent but rather are linearly coupled with
each other, and this hypothesis contradicted to experimental
findings~\cite{pollak}. Furthermore, recent observations of ES law
were made in the arrays of quantum dots \cite{Yakimov} and
artificially manufactured granular systems \cite{Hendrich1} where
the size of granules/periodicity in the dots arrangement was
controlled within the few percent accuracy. Thus the basic
assumption about the large size/distance dispersion does not apply
to all these systems, yet the dependence (\ref{hopping}) is
observed. Moreover, as was pointed out in
Refs.~[\onlinecite{pollak, Shklovskii04}] the capacitance disorder
alone can never lift the Coulomb blockade in a single grain
completely, and thus this mechanism for sure cannot explain the
finite density of states at the Fermi level, and as a consequence
it cannot explain dependence (\ref{hopping}) at low temperatures.

To address this problem properly, one notices that the insulating
matrix in granular conductors that is typically formed by the
amorphous oxide, may contain a deep tail of localized states due
to carrier traps.  The traps with energies lower than the Fermi
level are charged and induce the potential of the order of
$e^2/\kappa r$ on the closest granule, where $\kappa$ is the
dielectric constant of the insulator and $r$ is the distance from
the granule to the trap.  This compares with the Coulomb blockade
energies due to charging metallic granules during the transport
process. This mechanism was considered in
Ref.~[\onlinecite{Shklovskii04}]. If granules are very small,
surface effects also contribute to random changes in associated
Coulomb energies. And, finally, talking about the 2D granular
arrays and/or arrays of quantum dots, one expects that the source
of the induced random potential are imperfections of and the
charged defects in the substrate. Thus in granular conductors the
role of the finite density of impurity states is taken up by
random shifts in the granule Fermi levels due to all the above
reasons. We describe these shifts as the external random potential
$\mu_i$, where $i$ is the grain index, applied to each site of a
granular system.

Tunneling over the distances well exceeding the average granule
size in a dense granular array is a fundamental physical problem.
The virtual electron tunneling through a {\em single} granule
(quantum dot) was considered in Ref.~[\onlinecite{Averin}] where
two different mechanisms for a charge transport through a single
quantum dot in the Coulomb blockade regime were identified. The
first mechanism --  the so called elastic cotunneling -- transfers
the charge via tunneling of an electron through an intermediate
virtual state on the dot such that the electron comes out of the
dot with the same energy it came in. In the second mechanism --
inelastic cotunneling -- an electron that comes out of the dot has
a different energy from the one that came in. After such a
tunneling process an electron-hole excitation is left in the dot
which absorbs the energy difference. Note that both these
processes go via the classically inaccessible intermediate states,
i.e. both mechanisms occur as the charge transfer via a virtual
state. Inelastic cotunneling dominates the elastic one in the
region where the temperature or the applied voltage is larger than
$\sqrt{ E_0^c \delta }$.

Coulomb charge mapping technique, developed in our work, allows
for a simple and transparent analytic description of both elastic
and inelastic cotunneling processes offering thus a ground for a
comprehensive theory of the hopping transport in granular metals.
The advantage of our approach is that the higher order tunnelling
processes can be included in a straightforward way along with the
nearest neighbor hoppings. The obtained results for the hopping
conductivity dependencies as well as results concerning regular
arrays are summarized in the section below.

Our paper is organized as follows:  We summarize our results in
Sec.~\ref{summary} and introduce our model in details in
Sec.~\ref{modelsec}. In Sec.~\ref{gap} we find the corrections to
the Mott gap due to the weak intergranular coupling. In
Sec.~\ref{representation} we introduce the charge representation
for the model of a granular metal and discuss the physical
quantities in this representation in Sec.~\ref{quantities}. The
Mott gap for the periodic sample in the case of moderately large
tunneling conductance is found in Sec.~\ref{gap2}. In
Sec.~\ref{higher} we generalize our technique to include the high
order tunneling processes and apply this technique to find the
hopping conductivity in Sec.~\ref{disorder} and to study the Mott
transition in Sec.~\ref{Mott}. Our conclusions are presented in
Sec.~\ref{conclusions} while some technical details are relegated
into Appendix.

\section{Summary of the results and discussion}
\label{summary}

To treat the coherent multi-granule tunneling and to describe the
insulating state we have developed an approach based on the
mapping of the quantum model of granular metal onto the classical
gas of Coulomb charges in $d+1$ dimensions.  The Coulomb charges
of the effective classical model are connected by electron world
lines and form necklace-like loops.   First we apply our technique
to calculate the Mott gap $\Delta_M$ in the electron spectrum of a
periodic granular array and show that it is simply related to the
dielectric constant $\kappa_\tau$ of the classical charge gas in
the extra time direction. In the insulating regime where virtual
tunnellings on large distances are suppressed, the charge model
approach is reduced to consideration of the short loops,
representing the nearest neighbor hopping. Close to the
metal-insulator transition the larger loops that describe
tunneling on larger distances become important, and in the
metallic regime the loops of an infinite size appear. In this
picture, the Mott transition corresponds to the transition between
the state where large loops are energetically suppressed and the
state where loops of an infinite size do proliferate.  It is
important that despite the large loops representing tunneling over
large distances are energetically suppressed and do not contribute
to thermodynamics of the insulating state, they are essential for
transport and govern hopping conductivity in the arrays subject to
electrostatic disorder.

We will discuss the physics of the Mott insulator phase and
describe the metal to insulator transition in the context of the
periodic granular model.  The subsequent derivation of hopping
conductivity in the weakly coupled arrays subject to external
electrostatic disorder does not require special assumptions about
the spatial periodicity and holds both for regular and irregular
arrangements.

\subsection{ Periodic granular arrays }

An array of weakly coupled metallic grains, $g \ll 1$, is the
charge insulator at zero temperature. The electron transport is
blocked due to the presence of the Coulomb gap $\Delta_M$ in the
electron excitation spectrum.  In the limit of vanishingly small
intergranular tunneling conductances this gap is given by the
Coulomb charging energy of a single grain $\Delta_M = E^c_0$. At
finite tunneling conductance the Mott gap $\Delta_M$ is reduced
due to the virtual electron tunneling to neighboring grains.
Suppression of the Mott gap $\Delta_M$ due to such virtual
processes in the limit of weak coupling  can be obtained from the
perturbation theory and is given by
\begin{equation}
\label{Delta_M_per_0} \Delta_M = E^c_0 - {{2 z}\over{\pi}}\, g \,
E_{eh}\, \ln 2 , \;\;\;\; g z \ll 1,
\end{equation}
where $z$ is the coordination number  and $ E_{eh} $ is the energy
to create an electron-hole excitation in the system by removing an
electron from a given site and placing it on one of the
neighboring sites. In the case of diagonal Coulomb interaction
$E_{ij} = E_0^c \, \delta_{ij}$ the energy of an electron-hole
excitation $E_{eh}$   is simply twice the Coulomb charging energy
$E_0^c$.
   \begin{figure}[tbp] \hspace{-0.5cm}
   \includegraphics[width=2.8in]{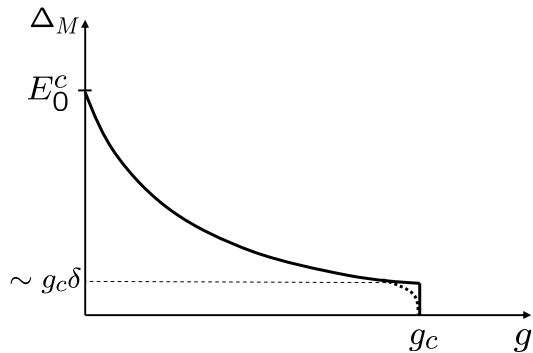}
   \caption{  Schematic dependence of the Mott gap on the tunneling
   conductance $g$ for a periodic granular array of the dimensionality $d > 2$.
   At zero tunneling conductance the Mott gap is given by the charging energy
   in a single grain and it suppressed exponentially with $g$  for $g z >1.$
   The Mott transition takes place at $\Delta_M\sim g\delta,$
   where $\delta$ is the mean energy level spacing in a single grain.
   The possible continuous  transition that cannot be ruled out at present time is shown by
   the dotted line.
    } \label{Mott_gap}
   \end{figure}
From Eq.~(\ref{Delta_M_per_0}) it follows that the Mott gap
$\Delta_M$ is reduced significantly at the values of the tunneling
conductance $g z \sim 1$, where at the same time the perturbation
theory becomes unapplicable.

In order to find the Mott gap $\Delta_M$ at larger tunneling
conductances, we make use of mapping of the original model onto
the model of classical charges and construct a mean field theory
considering the coordination number $z$ as a large parameter, $z
\gg 1$. Within this theory we obtain that the Mott gap $\Delta_M$
is exponentially suppressed with the tunneling conductance
 \begin{equation}
 \label{DelM}
 \Delta_M = c \, g\, E^c_0 \; e^{- \pi z g},   \;\;\;\;\; \Delta_M \gg
 g\delta\;\;\;\; g z \gtrsim 1,
 \end{equation}
where $c$ is the number of the order of unity. When deriving
equation~(\ref{DelM}), the nearest neighbor virtual electron
tunneling  only was taken into account. In the related effective
classical model this corresponds to accounting for the short
quadruple loops only. For this reason the result~(\ref{DelM})
holds only as long as the Mott gap $\Delta_M$ is larger than the
inverse electron escape time from the grain $\Gamma = g\delta$.
Solving the inequality $\Delta_M > g\delta$ together with the
expression for the Mott gap (\ref{DelM}) within the logarithmic
accuracy we obtain that the later can be used as long as  the
tunneling conductance $g$ is less than the critical conductance
given by the expression
 \begin{equation}
 \label{critical_conductance}
 g_c = {1\over { z \pi }} \; \ln (E^c_0 /\delta).
 \end{equation}
In the regime where the Mott gap $\Delta_M$ becomes of the order
of the inverse escape time of an electron from a grain $g\delta$
the large loops become important showing tendency to the formation
of the metallic state. This means that the value
(\ref{critical_conductance}) determines the boundary of stability
of the insulating state. At the same time the critical tunneling
conductance~(\ref{critical_conductance}) coincides with the
boundary of the stability of the metallic phase obtained in
Ref.~[\onlinecite{BLV03}]  meaning that the critical
value~(\ref{critical_conductance}) indeed marks the boundary
between the metallic and insulating phases at zero temperature.

Considerations leading to the result~(\ref{critical_conductance})
for the critical tunneling conductance $g_c$ are based on the
investigation of the behavior of the system from the insulating
and metallic sides implemented in the present paper and
Ref.~[\onlinecite{BLV03}] respectively. However, both these
considerations do not apply in the immediate vicinity of the phase
transition. At present the quantitative description of the Mott
transition is not available, in particular, the order of the
transition remains an open question. Two possible options are
illustrated on Fig.{\ref{Mott_gap}}.

Presented description of the Mott transition may not apply in one
and two dimensions where even the metallic state $g > g_c$ is
poorly defined due to divergent Althshuler-Aronov and localization
corrections to conductance at $T=0$.  The metal to insulator
transiton/crossover at higher temperatures, $T> g \delta$ in low
dimensional metallic arrays was considered in Refs.
[\onlinecite{Fazio,Jul'ka,Glazman}].

The temperature dependence of the conductivity of granular arrays
which are insulators at $T=0$ (i.e. with $g < g_c$), is given by
the activation law~\cite{Efetov02}
\begin{equation}
\label{activation}
 \sigma \sim \sigma_0 \; e^{ - \Delta_M /T },
\hspace{1cm}  T  \ll \Delta_M,
\end{equation}
with $\sigma_0$ being the high temperature conductivity as long as
the temperature is less than the Mott gap. Indeed, the finite
temperature conductivity is due to the electrons and holes that
are present in the system as real excitations . Their density is
given by the Gibbs distribution that results in the conductivity
dependence (\ref{activation}). Note that the
expression~(\ref{activation}) contains the renormalized Mott gap
$\Delta_M$ given by Eqs.~(\ref{Delta_M_per_0},\ref{DelM}).

It is important to note that at finite temperatures the continuous
transition  between the insulating and metallic phases, is, in
fact, rather a crossover  than the true phase transition.  Indeed,
the conductivity at nonzero temperature is finite in both phases,
besides the Mott transition does not break any symmetry of the
system. However, if the transition is discontinuous at $T=0$ it
has to remain discontinuous in some finite temperature interval.
In this case the Mott transition would be observable even at
nonzero temperature via the jumps of the physical quantities such
as conductivity.

The results presented above hold only for periodic arrays. Most of
realistic granular/quantum dots arrays contain disorder, in
particular, electrostatic disorder.

\subsection{ Granular arrays with electrostatic disorder }

Electrostatic disorder causes fluctuations in electrostatic energy
of granules and can thus lift the Coulomb blockade at some sites
of a granular sample.  This results, in its turn, in the finite
density of states at the Fermi level and in appearing the variable
range hopping as the dominating mechanism of conductivity. The
bare density of states induced by the random potential can be
substantially suppressed due to the presence of the long-rang
Coulomb interaction as it takes place in semiconductors where the
density of states (DOS) is given by the Efros-Shklovskii
expression~\cite{Efros,Shklovskii}
 \begin{equation}
 \label{ES_density_of_states}
 \nu_g (\varepsilon) \sim  ( \tilde \kappa / e^2)^d \; |\varepsilon|^{d-1},
 \end{equation}
with $e$ being the electron charge and $\tilde \kappa$ being the
dielectric constant.   This result
 was recently
 confirmed in Ref.\cite{Ioffe} (see also Ref.\cite{Pankov})  within
 the analytic approach of the locator approximation.

In the limit of weak coupling between the grains, $g \ll 1$, the
model of granular array can be described in terms of the classical
model that deals with the total charges on the grains only
 \begin{equation}
 \label{classical_model}
 H = \sum_i \mu_i n_i +\sum_{ij} n_i\; E^c_{ij} n_j,
 \end{equation}
where $\mu_i$ is the random external potential, $n_i$ is the
electron density on the site $i,$ and $E^c_{ij}$ is the Coulomb
interaction between the sites $i$ and $j.$ Taking into account the
asymptotic behavior of the Coulomb interaction matrix
  $E^c_{ij} \sim e^2/ r_{ij} \tilde \kappa ,$ where $r_{ij}$ is the distance between
the sites $i$ and $j$ and $\tilde \kappa$ is the effective
dielectric constant in a granular sample we see that the classical
model (\ref{classical_model}) is essentially equivalent to the one
studied by Efros and Shklovskii. Therefore their results must be
applicable to the model of a granular
array~(\ref{classical_model}) as well.

At the same time one expects that the DOS in an array of metallic
granules is larger than that in a semiconductor since each
metallic grain has a dense electron spectrum. Indeed, one has to
remember that there are many electron states in a grain that
correspond to the same grain charge while in the model of impurity
levels in semiconductors the charge is uniquely (up to the spin)
identified with the electron state. The energy of the unoccupied
state $\varepsilon_i$ in the model~(\ref{classical_model}) is by
definition  the energy of an electron placed on this state. In the
granular metal any state with energy larger than $\varepsilon_i$
will be also available for electron placement. Thus, in order to
translate ES result (\ref{ES_density_of_states}) to the density of
electron spectrum in granular metals one has to integrate the
dependence (\ref{ES_density_of_states}) by the energy
$\varepsilon$ and
 multiply it by the bare DOS in a single grain. As a result we obtain
  \begin{equation}
  \label{Granular_ES}
  \nu(\varepsilon) \sim \nu_0 \; ( |\varepsilon|  \tilde \kappa /
  \,e^2
  )^d,
  \end{equation}
where $\nu_0$ is the average DOS in a single grain (measured as
the number of states per energy). However, the above expression
cannot be used in the Mott argument for the hopping conductivity
where one needs to estimate the distance to the first available
state $r$ within the energy shell $\varepsilon$ via the relation
  \begin{equation}
  \label{est}
  r^d \; \int_0^\varepsilon d\varepsilon^\prime \nu_g(\varepsilon^\prime)  \sim
  1.
  \end{equation}

The problem with using the expression for DOS~(\ref{Granular_ES})
in Eq.~(\ref{est}) is that  expression~(\ref{Granular_ES}) assumes
strong correlations of the electron states in the space of the
array coordinates. Indeed, expression~(\ref{Granular_ES}) takes
into account the fact that if there is a state available for
placement of an electron with a given energy, then, typically {\it
on the same grain} there will be plenty of other states  available
for electron placement. But, for application to the hopping
conductivity we should not count different electron states that
belong to the same grain since it is enough to find at least one
to insure the transport. Thus, when finding DOS relevant for the
hopping transport the lowest energy states within the each grain
are to be count only. We arrive at the conclusion that while in
granular metals the electron DOS is modified according to
(\ref{Granular_ES}), in order to find the distance to the first
available state within a given energy shell via Eq.(\ref{est}),
even in granular metals, one has to use the ES expression for DOS
in its form (\ref{ES_density_of_states}). Similar considerations
were presented in Ref.\cite{Shklovskii04}. Following this
reference we will call the DOS that counts only lowest excited
states  in each grain and is relevant for the hopping conductivity
by the density of {\it ground } states and will denote by a
subscript  $g$ as in Eq.\ref{ES_density_of_states}.

  Apart from the knowledge of the spectrum of electron excitations, description
  of the hopping conductivity requires understanding the tunneling mechanism of an
  electron between the two distant grains with energies close to the Fermi level through
  an optimal chain of  other grains where the Coulomb blockade is
  present.  At  low temperatures such tunneling process
  can be realized by means of the virtual electron hopping (cotunneling).
  As we mentioned in Introduction one has to distinguish the
  elastic and inelastic cotunneling processes. The elastic cotunneling is the dominant
  mechanism for hopping conductivity at temperatures $T < \sqrt{E_0^c \delta}$,
  while at  larger temperatures  the electron transport
  goes via inelastic cotunneling processes.

  \subsubsection{Hopping via elastic cotunneling}
\label{sum_el_cot}

Considering the probability for an electron from  the site $i_0$
to tunnel to the site $i_{N+1}$ it is convenient to put the
Coulomb interaction energy at these sites to zero and count the
initial, $\xi_0$, and final, $\xi_{N+1}$, electron energies from
the Fermi level.  The presence of the electrostatic disorder on
the grains is modelled by the random potential $\mu_i,$ such that
the energy of the electron (hole) excitation on the site $i$ is
 \begin{equation}
 \label{eh}
 E_i^ {\pm}  = E_i^c  \pm \mu_i.
 \end{equation}
The probability $P_{el}$ of a tunneling process via the elastic
cotunneling can be easily found for the case of the diagonal
Coulomb interaction $ E_{ij}^c = E^c_i \, \delta_{ij}$:
\begin{equation}
\label{P_el_result}
 P_{el} =  w \, \bar g^{N+1} \, \left ( {{\bar \delta }\over {\pi  \bar E }}
 \right)^N  \delta(\xi_{N+1}-\xi_0 ),
 \end{equation}
where the factor
    $
    w = n(\xi_0)(1-n(\xi_{N+1}))
    $
takes into account the occupations $n(\xi_0)$ and $ n(\xi_{N+1}) $
of the initial and final states respectively, and the bar denotes
the geometrical average of the physical quantity along the
tunneling path. For example the average tunneling conductance
$\bar g$ is defined as
\begin{equation}
\ln \bar g = { 1\over {N+1} }\, \sum_{k=0}^{N}\, \ln g_k,
\end{equation}
where  the summation goes over the tunneling path and $g_k$ is the
tunneling conductance between the $k$-th and $(k+1)$-st grains.
The energy $\bar E$ is the geometrical average
$
\ln \bar E = { 1\over {N} }\, \sum_{k=1}^{N}\, \ln \tilde E_k,
$
of the following combination of electron and hole excitation
energies
\begin{equation}
\label{tilde_E}
 \tilde E_k =  2 \, \left ( 1/ E_k^+ +  1/E_k^-
\right)^{-1}.
\end{equation}
The presence of the delta function in  Eq.~(\ref{P_el_result})
reflects the fact that the tunneling process is elastic.

The result~(\ref{P_el_result}) for the tunneling probability
$P_{el}$ implies that on average the probability  falls of
exponentially with the distance $s$ along the path: $P_{el} \sim
e^{-2 s/\xi_{el}}$, where the dimensionless localization length
$\xi_{in}$ is given by
\begin{equation}
\label{localization}
 \xi_{el} =   {{ 2 }\over { \ln (\, \bar E \,
\pi / c \bar g \, \bar \delta ) }}.
\end{equation}
The numerical constant $c$ in the above expression is equal to one
for the model of diagonal Coulomb interaction, $ E_{ij}^c = E^c_i
\, \delta_{ij}$.   We show below that the inclusion  of the
off-diagonal part of the Coulomb interaction results in the
renormalization of the constant to some value $ 0.5 \lesssim  c <
1$.

In a full analogy with the hopping conductivity in
semiconductors~\cite{Shklovskii} in the case of granular metals
inelastic processes are required to allow an electron to tunnel to
a state with higher energy. In granular metals such processes
occurs via phonons as well as via the inelastic collisions with
other electrons; these proceses are not reflected in the formula
(\ref{P_el_result}). Applying the conventional
Mott-Efros-Shklovskii arguments, i.e. optimizing the full hopping
probability $\propto\exp[-(2s/\xi_{el}) -(e^2/\tilde\kappa Tas)]$,
one obtains Eq.~(\ref{hopping}) for the hopping conductivity with
the characteristic temperature
\begin{equation}
\label{T0} T_{\circ} \sim e^2 / { a  \tilde \kappa} \, \xi_{el},
\end{equation}
where $\tilde \kappa$ is the effective dielectric constant of a
granular sample, $a$ is the average grain size, and $\xi_{el}$
given by  Eq.~({\ref{localization}}) (when deriving (\ref{T0}) we
considered the tunneling path as nearly straight).

 In the presence of the strong electric field ${\cal E}$, the direct
 application of the consideration of Ref.~[\onlinecite{Shklovskii73}] for
 the case of low temperatures
 $
 {{ T }\over { e\xi_{el} a }} \ll {\cal E} \ll {{ \sqrt{ \delta  E_{\circ}^c  }  }\over { e a }}
 $
 gives the nonlinear current dependence
 \begin{equation}
 j \sim j_0 \; e^{-({\cal E}_{\circ}/{\cal E})^{1/2}},
 \end{equation}
 where the characteristic electric field ${\cal E}_{\circ}$ is given by
 the expression
 \begin{equation}
 {\cal E}_{\circ} \sim T_{\circ} / e \,a  \xi_{el} .
\end{equation}
 The results presented in this subsection are valid as long as the contribution
 of the inelastic cotunneling to the hopping conductivity can be
 neglected. This is the case at low temperatures and electric fields
 $T,{\cal E} e a \ll \sqrt{\delta E_0^c}.$ Below we present our results for the
 opposite case where inelastic processes give the main contribution to hopping conductivity.

\subsubsection{Hopping via inelastic cotunneling}

Inelastic cotunneling is the process where the charge is
transferred by means of  different electrons on each elementary
hop. During such process the energy brought to a grain by incoming
electron differs from the energy taken by the outcoming one.

As in the case of elastic cotunneling we assume that the electron
tunnels from the grain $i_0$ with energy $\xi_0$ to the grain
$i_{N+1}$ with energy $\xi_{N+1}.$  For the probability of such a
tunneling process through a chain of grains in the tunneling path
we find the general expression \cite{footenote2}
   \begin{equation}
   \label{result_in1}
   P_{in} = { w \over { 4\pi T }} {{   \, \bar g^{N+1}  } \over {\pi^{N+1}
   }}
   \left[ {{ 4 \pi  T }\over {  {\bar E } }} \right]^{2N}
   { { |\Gamma(N+{{ i \Delta}\over {2\pi T}} ) |^2 }\over {\Gamma(2N)
   }} e^{-{{\Delta}\over {2T}}},
   \end{equation}
where $\Gamma(x)$ is the Gamma function and
$\Delta=\xi_{N+1}-\xi_0$  is the energy difference between final
and initial states. The appearance of the factor $e^{-\Delta/2T}$
is consistent with the detailed balance principle. Indeed, at
finite temperatures an electron can tunnel in both ways with
either increase or decrease of its energy. According to the
detailed balance principle the ratio of such probabilities must be
$e^{\Delta/T},$ that is indeed the case for the function
$P_{in}$~(\ref{result_in1}), since apart from the factor
$e^{-\Delta/2T}$ the rest of the equation is even in $\Delta.$

To obtain the expression for the hopping conductivity  one has to
optimize Eq.~(\ref{result_in1}) with respect to the hopping
distance $N$ under the constraint
   \begin{equation}
   \label{constraint1}
   N \, a\, \tilde \kappa \, \Delta / e^2  \sim 1,
   \end{equation}
following from the ES expression for the density of ground
states~(\ref{ES_density_of_states}). Optimization of
Eq.~(\ref{result_in1}) is a bit more involved procedure than the
standard derivation of Mott-Efros-Shklovskii law based on the
Gibbs energy distribution function (see Appendix), however it
leads to the essentially the same result, i.e. the ES law with
   \begin{equation}
   T_0(T) \sim  e^2 /a \tilde \kappa \,  \xi_{in}(T),
   \end{equation}
with the dimensionless localization length $\xi_{in}(T)$ being
weakly temperature dependent
   \begin{equation}
   \label{inelastic_loc_len}
   \xi_{in}(T) = { 2 \over {\ln [ \,\bar E^2 / 16 c \pi T^2  \bar g   \,] }},
   \end{equation}
where the coefficient $c$ equals one for a model with diagonal
Coulomb interaction matrix. As in the case of the inelastic
cotunneling, the long range interaction result in the reduction of
the coefficient to some $1/4 \lesssim c < 1$.

At zero temperature Eq.~(\ref{result_in1}) for inelastic
probability $P_{in}$ for $\Delta <0$  is simplified to
   \begin{equation}
   \label{inel}
   P_{in}(T=0) =  {{ w\, 2^{2N} \pi } \over {  (2N-1) !}}
    {{ |\Delta|^{2N-1} }\over {  {\bar E  }^{2N} }} \left( {{   \, \bar g  } \over {\pi
   }}\right)^{N+1},
   \end{equation}
while for $\Delta >0 $ Eq.~(\ref{result_in1}) gives zero since the
process with increase of the energy of the tunneling electron is
prohibited at zero temperature.

To obtain the hopping conductivity in the regime of strong
electric field $\cal E$ and low temperatures, following
Ref.~[\onlinecite{Shklovskii73}], we use the
condition~(\ref{constraint1}) that defines the distance to the
first available electron site within the energy shell $\Delta$
together with equation
   \begin{equation}
   \label{Electric_Filed}
   e {\cal E} r \sim \Delta,
   \end{equation}
that relates $\Delta$ with the electric filed ${\cal E}$ and the
distance between the initial and final tunneling sites $ r \sim
Na.$ Together Eqs.~(\ref{constraint1}) and (\ref{Electric_Filed})
define
   $
   \Delta \sim \sqrt{ {\cal E } e^3 / \tilde \kappa   }
   $
and $ N \sim \sqrt{ e / \tilde\kappa {\cal E} a^2  } $ that being
inserted into Eq.~(\ref{inel}) result in the expression for the
current
    \begin{subequations}
    \begin{equation}
     j \sim j_0 \; e^{-({\cal E}_0/ {\cal E})^{1/2}}.
    \end{equation}
where the characteristic electric field ${\cal E}_0$ is a weak
function of the applied filed ${\cal E}$
   \begin{equation}
   {\cal E}_0({\cal E}  )  \sim {{ e }\over {\tilde \kappa \, a^2 }} \;
   \ln^2[\bar E^2 / e^2 {\cal E}^2 a^2 \bar g ].
   \end{equation}
\end{subequations}

\section{The Model}
\label{modelsec}

Above presented results were derived within the following model:
We consider a $d$-dimensional array of metallic grains. The
electron motion within each grain is diffusive and the grains are
assumed to be weakly coupled such that the  intergranular
tunneling conductance $g$ is much less than the intragranule one.
The system is described by the Hamiltonian
\begin{subequations}
\label{model}
\begin{equation}
\hat{H}=\hat{H}_{0}+\hat{H}_{c}+\sum_{ij} \,t_{k_i k_j}^{ij}  \,
\hat{\psi}^{\dagger }_{i}( k_i)  \;  \hat{\psi}_{j}( k_j ),
\label{hamiltonian}
\end{equation}
where $t_{k_i k_j}^{ij}$ is the matrix element representing the
electron tunneling amplitude from the state $k_i$ of the grain $i$
to the state $k_j$ of the neighboring grain $j.$ The Hamiltonian
$\hat{H}_{0}$ in the r.~h.~s. of Eq.~(\ref{hamiltonian}) describes
noninteracting isolated disordered grains. The term $\hat{H}_{c}$
includes the electron Coulomb interaction and the local external
electrostatic potential $\mu_i$ on each grain
\begin{equation}
\hat{H}_{c} = \sum_{ij} \, \hat{n}_{i} \,E^c_{ij} \, \hat{n}_{j} +
\sum_i \mu_i \hat{n}_i , \label{Coulomb}
\end{equation}
\end{subequations}
where $E^c_{ij}$ is the matrix of Coulomb interaction and
$\hat{n}_{i}$ is the operator of electron number in the $i-$th
grain. The Coulomb interaction matrix is related to the
capacitance matrix in the usual way
\begin{equation}
E^c_{ij} = ( e^2/2 ) \; C^{-1}_{ij}.
\end{equation}
As we discussed in Introduction the random potential $\mu_i$ that
models the electrostatic disorder removes  the Coulomb blockade
from some grains.

We let the tunneling elements be random Gaussian variables defined
by their correlators:
\begin{equation}
\label{def_t^2}
 \langle \; t_{k_1,k_2}^{*ij} \, t^{ij}_{k_1^\prime,k_2^\prime}
\;\rangle = t^2_{ij} \; \delta_{k_1 k_1^\prime} \; \delta_{k_2
k_2^\prime },
\end{equation}
where $i,j$ are the coordinates of the nearest neighbor grains.
The intergranular conductance between the grains $i$ and $j$ is
related to average matrix elements as
\begin{equation}
g_{ij} = 2\pi  \nu_i \nu_j   t_{ij}^2
\end{equation}
where $\nu_i$ is the density of states in the grain $i$.  The
conductance $g_{ij}$ is defined per one spin component, such that,
for example,  the high temperature ( Drude ) conductivity of a
periodic granular sample is $\sigma = 2 e^2 g a^{2-d},$ with $a$
being the size of the grain.  In the following we will consider
both cases of periodic ( in particular, it means that $\mu_i = 0,
\; \nu_i=\nu_0, \; g_{ij} = g $) and irregular arrays beginning
with the former case.

\section{Mott gap at small tunneling conductances} \label{gap}

In the limit of small tunnelling conductances, $g \ll 1$, and zero
temperature a {\it periodic} granular array is a charge insulator.
The key characteristic of the insulating state is the Mott gap
$\Delta_M$ in the electron excitation spectrum that can be defined
as
 \begin{equation}
 \Delta_M =  E_{N=1} - \mu - E_{N=0}, \label{Delta}
 \end{equation}
where $\mu$ is the chemical potential, $ E_{N=0}$ is the ground
state energy of the charge neutral array and $ E_{N=1} $ is the
minimal energy of the the system with an extra electron added to
the neutral state.  In the limit of vanishing tunneling
conductance, $g \to 0$, the Mott gap $\Delta_M$ coincides with the
Coulomb energy $E^c_0  \equiv E^c_{i,i}$ in a single grain
 \begin{equation}
 \Delta_{M\, g \to 0} =E^c_0.
 \end{equation}

At finite intergranular coupling the Mott gap is reduced as a
result of the virtual presence of an electron on the neighboring
grains. In the case of weak coupling,  $g \ll 1$, the resulting
suppression of the Mott gap can be easily calculated within the
perturbation theory. The first non vanishing correction appears
only in the second order in tunneling matrix elements
 and  can be found with the help of the textbook formula~\cite{Landau}
 \begin{equation}
 \Delta E = - \sum_k {{ |V_{k,0}|^2 }\over { E_k-E_0}},  \label{second_order}
 \end{equation}
where $\Delta E$ is the energy correction to the ground state and
the matrix elements of the perturbation $V$ are taken between the
ground state $0$ and excited states $k$.

\begin{figure}[tbp]
\hspace{-0.5cm}
\includegraphics[width=3.4in]{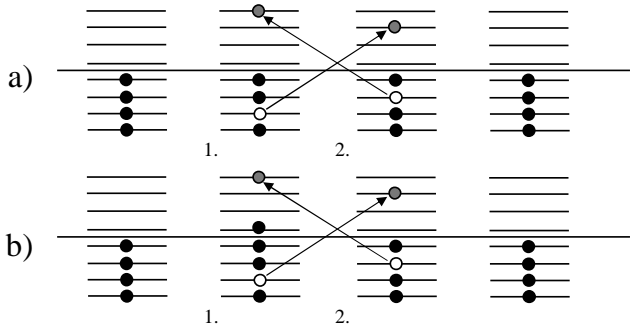}
\caption{ This figure illustrates the virtual electron tunneling
processes between  neighboring grains that contribute to the
reduction of the Mott gap. The tunneling processes shown on figure
a) determine the correction to the ground state energy of the
charge neutral state $E_{N=0}.$ The correction to the energy of
the ground state with one extra electron (placed on grain 1)
$E_{N=1}$ is illustrated in figure b). } \label{Mott_correction}
\end{figure}

Correction due to finite intergranular coupling should be included
in both terms $E_{N=1}$ and $ E_{N=0}$ in Eq.~(\ref{Delta}). The
ground state with one extra electron $N=1$ in the limit of zero
coupling is degenerate since the extra electron can be present on
any grain. Further, we will assume that it occupies  grain 1 as it
shown in Fig.~\ref{Mott_correction} b. Since we want to find the
difference between the corrections to $N=1$ and $N=0$ ground
states, it is clear that within the second order perturbation
theory in tunneling elements $t_{k_1,k_2}$ only electron hops
between grain 1 and its neighbors result in nonzero contribution.
All other possible hops give rise to equal corrections to the
energies $E_{N=1}$ and $E_{N=0}$ that are mutually cancelled in
Eq.(\ref{Delta}). Further, since all the neighbors of grain 1 are
equivalent, we will consider only electron hops between  grain 1
and one of its neighbors that we denote as grain 2. Contribution
of the hops between  grain 1 and all its other neighbors will
simply result in the factor $z$ (coordination number ) in the
final answer.


Now let us consider the correction to the ground state energy
$E_{N=0}$. In this case the matrix elements $V_{k,0}$ correspond
to the processes of electron tunneling between the neutral
neighboring grains. Consider the tunneling process from grain 1 to
grain 2 shown in Fig.~\ref{Mott_correction}a: The excitation
energy of this process is  $\varepsilon_{k_1} +\varepsilon_{k_2} +
E_{eh},$ where $\varepsilon_{k_2}$ is the bare (with no Coulomb
energy included) energy of an electron excitation in grain 2,
$\varepsilon_{k_1}$ is the bare energy of a hole excitation in
grain 1, and $E_{eh}$ is the Coulomb part of the energy of the
electron-hole excitation $E_{eh}=E^c_{11} +E^c_{22}-2E^c_{12}$
that for the periodic case under  consideration is reduced to
\begin{equation}
\label{electron_hole}
 E_{eh}=2E^c_0-2 E^c_{12}.
\end{equation}
 The energy correction corresponding to such process is
\begin{equation}
\label{Delta0} - \Delta E_{N=0} =  2 \sum_{k_1,k_2} \, {{ |
t_{k_1,k_2} |^2 }\over{\varepsilon_{k_1}+\varepsilon_{k_2} +
E_{eh} }},
\end{equation}
where the factor 2 takes into account the similar process of
electron hopping from grain 2 to grain 1.

Analogously we find the correction to the energy $E_{N=1} $ as
illustrated on Fig.~\ref{Mott_correction}b:
  The excitation energy of the process
of electron tunneling from grain 2 to grain 1   is
$\varepsilon_{k1}+\varepsilon_{k_2} + 3
E^c_{11}+E^c_{22}-4E^c_{12} = \varepsilon_{k_1}+\varepsilon_{k_2}
+ 2 E_{eh} $, while the excitation energy of the process of
electron tunnelling from grain 1 to grain 2 is
$\varepsilon_{k_1}+\varepsilon_{k_2} + E^c_{11}-E^c_{22} =
\varepsilon_{k_1}+\varepsilon_{k_2} .$ The corresponding
correction to the energy $E_{N=1}$ reads
\begin{equation}
\label{Delta1} -  \Delta E_{N=1} =  \sum_{k_1,k_2} \, {{ |
t_{k_1,k_2} |^2 }\over{\varepsilon_{k_1}+\varepsilon_{k_2} +
2E_{eh} }} + {{ | t_{k_1,k_2} |^2
}\over{\varepsilon_{k_1}+\varepsilon_{k_2} }}.
\end{equation}
One can see that corrections~(\ref{Delta0}, \ref{Delta1}) are
ultraviolet divergent. However, their difference must be finite
since it represents the measurable quantity - correction to the
Mott gap. It is, indeed, the case: Subtracting Eq.~(\ref{Delta0})
from Eq.(\ref{Delta1}), presenting summation over states as
integrals $\varepsilon_{k_1(k_2)} \to \varepsilon_{1(2)} $ and
introducing further the variable $\varepsilon =
\varepsilon_1+\varepsilon_2$ we obtain
\begin{equation}
\Delta_M -E^c_0 = z \nu_0^2 t^2 \int_0^\infty \varepsilon \,
d\varepsilon \left[ {1\over {\varepsilon+2 E_{eh} }} - {2\over
{\varepsilon+E_{eh} } } +{1\over{\varepsilon}} \right], \nonumber
\end{equation}
where $\nu_0$ is the density of states in a single grain and $t^2$
defined by Eq.(\ref{def_t^2}) appears from the averaging over
matrix elements. Taking the integral in the above expression we
obtain the result~(\ref{Delta_M_per_0}).

In the derivation of the reduction of the Mott gap  presented in
this section we neglected the fact that an extra electron added to
the neutral system in the presence of finite intergranular
coupling can move diffusively over the sample. Contributions
corresponding to such processes, as we will see below,  are
suppressed by an extra small factor $\delta/E_0^c \ll 1$ and,
thus, can be neglected.

\section{Phase and charge representations} \label{representation}

The perturbation theory presented in Sec.~\ref{gap} is limited to
the regime of low tunneling conductance $g \ll 1$. Moreover, the
high order tunneling processes that are very important for
understanding the physics of Mott transition and hopping
conductivity cannot be included in this way. For this reason one
has to seek for a  convenient approach applicable at intermediate
tunneling conductances and capable of description of the effects
of high order tunnelling processes.  In this view, a useful tool
in approaching this problem is the phase functional approach
\cite{AES} that allows further mapping of the system to the
Coulomb gas of charged particles in d+1 dimensions
\cite{Schmid,Zaikin}.

\subsection{Phase representation}

The phase functional approach is based on the decoupling of the
Coulomb interaction term ~(\ref{Coulomb}) in the Hamiltonian
(\ref{hamiltonian}) by means of the the potential field
 $V(\tau)$ that is further absorbed into the gauge
 transformation of the fermionic fields
 \begin{equation}
 \label{gauge}
\psi_i(\tau) \to \psi_i(\tau) \, e^{-i\,\phi_i(\tau)}, \;\;\;\;\;
\phi_i(\tau) = \int_{-\infty}^\tau V_i(\tau^\prime) d\tau^\prime .
\end{equation}

Since the action of an isolated grain is gauge invariant the
phases $\phi_i(\tau)$ enter the whole lagrangian of the system
only through the tunnelling matrix elements
\begin{equation}
\label{tilde_t} t_{ij} \to t_{ij} \, e^{i\phi_{ij}(\tau)},
\end{equation}
where $\phi_{ij}(\tau) = \phi_i(\tau) - \phi_j(\tau)$ is the phase
difference of the $i-$th and $j-$th grains.

A simple description in terms of phase variables can be obtained
via integrating out the fermionic fields and expanding the
resulting rather complicated action to the lowest non vanishing
order in tunneling elements. As a result one obtains the
Ambegaokar, Eckern, Sch\"{o}n (AES) action~\cite{AES}
\begin{subequations}
\label{AES}
\begin{equation}
\label{AES_1} S_{AES}=-{1\over {2 e^2}}\sum_{ij} \int d\tau \,
(\dot \phi_i +i\mu_i) \, C_{ij} \, ( \dot \phi_j +i \mu_j)
+S_t[\phi],
\end{equation}
where the tunneling term $S_t$ at zero temperature is
\begin{equation}
\label{Z} S_t[\phi] =  {1 \over {2 \pi }} \sum\limits_{<ij>}
g_{ij} \int\limits_{-\infty}^{+\infty} \, d\tau_1 \, d\tau_2
 {{ e^{i\phi_{ij}(\tau_1) - i\phi_{ij}(\tau_2)} }\over
{ (\tau_1-\tau_2)^2 }},
\end{equation}
\end{subequations}
where the angular brackets mean summation over the nearest
neighbors. The higher order terms in the expansion of the total
action in the tunneling matrix elements  cannot be always
neglected and, in general, must be taken into account along with
the nearest neighbor action (\ref{AES}). However, before
considering these more complicated higher order terms, it is
instructive to   analyze,  first, the action (\ref{AES}) in
details. Moreover, as we will show, even at $T=0$ there is a
regime  where the AES action is applicable.  We, thus, proceed to
the description of the charge mapping  beginning with the AES
action, and later, in Sec.~\ref{higher}, we will show how the
higher order tunneling terms can be included into the mapping.
Also, for simplicity, we will first consider the case of the
regular periodic array.

\subsection{Charge representation} \label{b}

 Proceeding to the mapping of the action~(\ref{AES}) onto the model of the classical Coulomb
 gas and
 using the method of Ref.~[\onlinecite{Schmid}], we first expand the partition
 function $Z$ to all orders in the tunneling part of the action $S_t$
 \begin{equation}
  Z = Z_0 \, \sum_{N = 1}^{\infty} \, \langle \; S_t^N [\phi] \;
  \rangle \, / \, N ! \, . \label{NN_Partition_Function}
 \end{equation}
 Here $Z_0$ is the partition function of a system of isolated
 grains and the angle brackets assume averaging with respect to the
 Coulomb phase action that for the case $\mu_i=0$ under
 consideration reads
 \begin{equation}
 S_c = - {1\over {2 e^2}}\sum_{ij} \int d\tau \, \dot \phi_i \,
 C_{ij} \, \dot\phi_j.
 \end{equation}
 On the next step, for each term in the expansion~(\ref{NN_Partition_Function})
 we write the $N$-th power of the tunnelling action $S_t$ explicitly as the product
\begin{subequations}
  \begin{equation}
  \label{product}
  S_t^N[\phi] = \prod_{n=1}^N \, S_t^{(n)},
  \end{equation}
  where the index $n$ of the tunneling action $S_t$ reminds that the time
  integrations in each term in the above product are independent
  \begin{equation}
  \label{S_n}
  S_t^{(n)} = {{g }\over {2 \pi}} \sum\limits_{<ij>}
  \int\limits_{-\infty}^{+\infty} \, {{d\tau_n \,
  d\tau_n^\prime}\over { (\tau_n-\tau_n^\prime)^2 }}
   \; e^{i\phi_{ij} (\tau_n) - i\phi_{ij}(\tau_n^\prime)  }.
  \end{equation}
\end{subequations}
 Writing the powers of the tunneling action as products (\ref{product})  allows to
 implement the integrations over the phase fields $\phi_i(\tau)$ exactly for each term in the
 expansion~(\ref{NN_Partition_Function}) with the help of
 the formula
 \begin{equation}
 \label{U_}
 \left\langle e^{i \sum_n \phi_i(\tau_n) e_n  } \right\rangle =
 e^{-U^c},
 \end{equation}
 where $U^c$  given by the expression
\begin{equation}
 U^c={1\over 2} \sum_{n_1,n_2} E^c_{i_{n_1} i_{n_2}}
 |\tau_{n_1}-\tau_{n_2}| \, e_{n_1}\, e_{n_2},  \label{interaction}
 \end{equation}
 can be viewed as a Coulomb energy of a system of
 classical charges in a space with an extra time
 dimension~\cite{temperature}.
\begin{figure}[tbp]
\hspace{-0.5cm}
\includegraphics[width=3.0in]{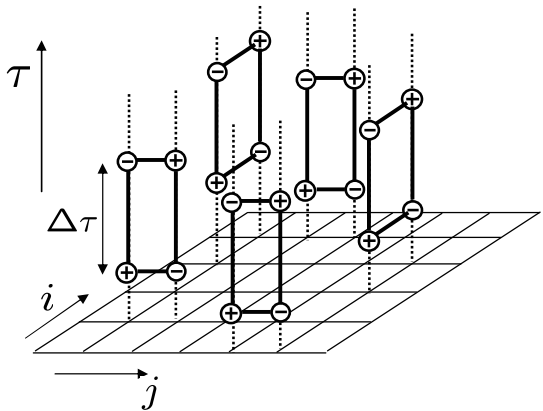}
\caption{ Illustration of the mapping of a quantum granular array
model in the nearest neighbor hopping approximation onto the
classical model of gas of quadruples in $d+1$ dimensions. }
\label{quadruple}
\end{figure}
The classical charges  $e_n=\pm 1$ interact via the 1d Coulomb
potential along the time direction while the strength
 of the interaction is given by the ``real" Coulomb  energy of
 original quantum electrons $E^c_{ij}.$

 From Eq.~(\ref{S_n}) we see that each tunneling term  $S_t^{(n)}[\phi]$
 contains  the sum of four phase fields in the exponent
 \begin{equation}
i\phi_{i} (\tau_n) -i\phi_j(\tau_n) - i\phi_{i}(\tau_n^\prime) +
i\phi_{j}(\tau_n^\prime).
 \end{equation}
 Each of these fields, according to  Eqs.~(\ref{U_},\ref{interaction}), gives rise to a classical charge
 located at coordinates  corresponding to the arguments of the phase fields.
 This means that the  classical charges appear as quadruples that
 occupy only  the neighboring sites as it is shown in Fig.~\ref{quadruple}.

 The time dependent kernel in Eq.(\ref{S_n}) can be written in the
 form
 \begin{equation}
   {{ g  } \over { 2 \pi } }  \; {1\over {(\tau_n-\tau^\prime_n )^2}} = e^{ - {\cal U}^q_n(\tau_n-\tau^\prime_n)
   },
 \end{equation}
 such that ${\cal U}^q_n$ defines the "internal" energy of $n-$th quadruple
 \begin{equation}
 \label{qudruple_energy}
 {\cal U}^q_n  = \ln [  2 \pi ( \Delta\tau_n )^2 /g ],
 \end{equation}
with  $\Delta\tau_n$ being the size of the $n-$th quadruple in the
$\tau-$ direction as shown in Fig.~\ref{quadruple}.

Thus, the partition function, Eq.~(\ref{NN_Partition_Function}),
of the quantum problem under consideration in the nearest neighbor
electron tunneling approximation can be presented as a partition
function of classical charges that appear in quadruples  all
charges being subject to the Coulomb
interaction~(\ref{interaction}). The total energy of the system of
N quadruples can be written as
\begin{equation}
U^q = \sum_{n=1}^N {\cal U}^q_n + U_{4N}^c \label{Charge1},
\end{equation}
where ${\cal U}^{q}_n$ is the internal energy of the $n-$th
quadruple and $U_{4N}^c$ is the  Coulomb energy of 4N charges that
form N quadruples defined by Eq.~(\ref{interaction}). We notice
that different quadruples interact with each other only via the
Coulomb part of the energy $U^c.$

In the following sections we will consider the thermodynamic
properties of the classical model and set correspondence between
the  parameters of the original quantum model and the effective
classical one.

\section{ Thermodynamics of the classical model in the limit of low quadruple density }
\label{quantities}

The thermodynamic potential $\Omega$ of the classical
model~(\ref{Charge1}) can be found in the limit of the low
quadruple density that, as we will show below, corresponds to the
case of the low tunneling conductance $g \ll 1$ in  the quantum
model. In this limit one can neglect the mutual interaction of
deferent quadruples considering them as free independent objects
and the thermodynamic potential $\Omega$ of the whole system is
reduced to the sum of the free energy  ${\cal F}_0$ of a single
quadruple multiplied by the number of the quadruples $N$ and the
entropy $S$ of the quadruple configuration space
 \begin{equation}
 \Omega = N {\cal F}_0 - S.
 \end{equation}
 The entropy of $N$ independent quadruples is given by the expression
\begin{equation}
\label{S}
 S = N \ln V - \ln N!,
 \end{equation}
 where $V =LM$ is the  volume of the $d+1$ dimensional system with L being
 the size of the system in  $\tau-$ direction  and $M$ being the total number of the
 sites (grains) in the array. The term $\ln N!$ in Eq.~(\ref{S}) appears
 from the term $1/N!$ in the expansion~(\ref{NN_Partition_Function}). In the classical
 statistical mechanics such  term appears due to equivalence of the classical configurations that
 differ only by mutual permutations of particles. Thus, in the
 thermodynamic limit $N\to\infty$  for the thermodynamic potential $\Omega$ in the case of low quadruple density
 we obtain the expression
 \begin{equation}
 \Omega=N {\cal F}_0 + N \ln \rho -N ,   \label{thermo_pot}
 \end{equation}
 where $\rho = N/V$ is the density of quadruples in $d+1$ dimensions.

 The partition function of a single quadruple is given by the
 expression
 \begin{equation}
 \label{z0}
 {\cal Z}_0 = {g\over {\pi}}\,
 \int_0^\infty {{ d \tau}\over {\tau^2 }}  \, e^{ - E_{eh}\, \tau
 },
 \end{equation}
 where the energy of the electron-hole excitation is defined by
 Eq.(\ref{electron_hole}). Equation~(\ref{z0}) is divergent on low integration limit and in order to regularize
 it we introduce the time cut-of $\lambda$  obtaining
 $ {\cal Z}_0 =  g / \pi  \lambda $ so that the free energy of a single quadruple ${\cal F}_0 = - \ln {\cal Z}_0$ becomes
  \begin{equation}
  \label{F_0}
  {\cal F}_0= -\ln(g /\pi  \lambda ).
  \end{equation}

  \subsection{ Density of classical charges }

 The thermodynamic potential given by Eqs.(\ref{thermo_pot},\ref{F_0})
determines  all the thermodynamic quantities of the classical
model in the limit of low quadruple density. In particular, the
density itself can be obtained via minimization of $\Omega$ with
respect to the
 number of quadruples $N$ resulting in
 \begin{equation}
  \rho =  g /  \pi \lambda.  \label{rho}
 \end{equation}
 We would like to note that the obtained density coincides with
the single quadruple partition function $ \rho= {\cal Z}_0. $
 The result for the density $\rho$~(\ref{rho}) can be
better understood considering a discrete time axis. In this case
the cut-of distance $\lambda$ would be the distance between the
neighboring  sites in $\tau-$ direction and the dimensionless
density according to Eq.~(\ref{rho}) would be $ g / \pi .$ Thus,
it becomes clear that the limit of low conductance $g \ll 1$
corresponds to the limit of the low occupation number of classical
charges.

\subsection{ Relation between the Mott gap and dielectric constant of classical gas }

 In order to make the developed mapping useful one has to express  the
 physical observables of the original quantum model in terms of
 the corresponding thermodynamic quantities of the  classical model.
 Below we show that the key characteristic of the
 insulating state - the Mott gap $\Delta_M$ is related to the
 dielectric constant $\kappa_\tau$ of the effective classical model
 measured in the direction of $\tau-$ axis as
 \begin{equation}
 \Delta_M =  E_0^c / \kappa_\tau.         \label{delta_kappa}
 \end{equation}

To prove relation~(\ref{delta_kappa}) we first note that the gap
$\Delta_M$ in the electron spectrum determines the asymptotic
behavior ( $\tau\to\infty$ )  of the single electron Green
function
 \begin{equation}
 G_{ii}(\tau) \sim e^{-\Delta_M \, |\tau|}.
 \end{equation}
At the same time the asymptotic behavior of the Green function
$G_{ii}(\tau)$ with exponential accuracy coincides with the
asymptotic behavior of the phase correlation function
 \begin{equation}
 \label{phase}
 G_{ii}(\tau_1-\tau_2) \sim \Phi(\tau_1 - \tau_2) \equiv
  \langle \, e^{i\phi_i(\tau_1)-i\phi_i(\tau_2)} \,
 \rangle,
 \end{equation}
where the angular brackets stand for averaging with respect to the
phase functional~(\ref{AES}), written  explicitly
  \begin{equation}
  \label{phi_phi}
  \Phi(\tau_1 - \tau_2) =  Z^{-1} \, \int D[\phi] \; e^{i\phi_i(\tau_1)-i\phi_i(\tau_2)} \, e^{S[\phi]}
  .
  \end{equation}
   All the considerations that was previously used in  order to derive the effective model (\ref{Charge1})
   can be applied to the correlation function (\ref{phi_phi}) as well.
   Repeating all the necessary steps we notice that  phases $\phi_i(\tau_1)$ and $\phi_i(\tau_2)$
   that enter the exponent in   Eq.(\ref{phi_phi}) result in appearance of two external charges
   of opposite signs located on the site i at time coordinates  $\tau_1$ and $\tau_2.$ Further, we see
   that the phase correlation function $\Phi$  is related to the thermodynamic potential of the
   system with two external charges $ \Omega_2(\tau_1-\tau_2) $  as
   \begin{equation}
   \Phi(\tau) = e^{-\Omega_2(\tau)}.
   \end{equation}
   The function $\Omega_2(\tau)$ is normalized such that
   $\Omega_2(0)=0.$ The function $\Omega_2$ is nothing but effective
   interaction of two external charges in the presence of quadruples.
    The initial electrostatic interaction is reduced in the presence of quadruples and the
   asymptotic behavior of the function $\Omega_2$ takes the form
   \begin{equation}
   \label{Omega_2}
   \Omega_2(\tau) = |\tau| \, E^c_0  / \kappa_\tau , \;\;\;\;\;\;\;
   \tau\to\infty,
   \end{equation}
   where $\kappa_\tau$ is the dielectric constant.  This proves the
   relation between the Mott gap and dielectric constant of the classical Coulomb
   gas in $\tau-$ direction (\ref{delta_kappa} ).

\subsection{ Mott gap in the limit of $g z \ll 1$  }

   Finally, in order to  better understand the relationship
   between the original quantum model and the effective classical one
   we will rederive the result~(\ref{Delta_M_per_0}) for the Mott
   gap in the case of small tunneling conductance $g z \ll 1$
   within the framework of the effective classical model.

\begin{figure}[tbp] \hspace{-0.5cm}
\includegraphics[width=3.0in]{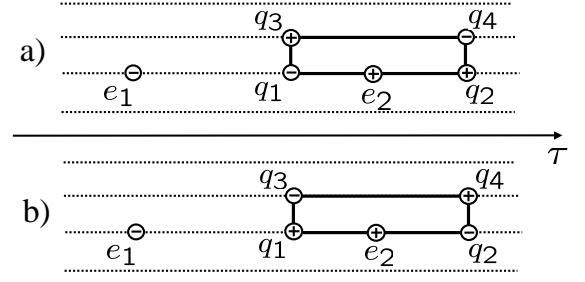}
\caption{ An electric filed that charge $e_2$  feels consists of
the field created by the charge $e_1$ and the fields created by
the charges of the quadruple $q_1,... q_4.$ }
\label{quadruple_screening}
\end{figure}

   Let us consider two charges $e_1$ and $e_2$ ( see Fig.~\ref{quadruple_screening} )
   of the opposite signs  located on the site $i$ at time coordinates $\tau_1$ and $\tau_2$
   respectively. In the absence of surrounding quadruples their interaction energy
   would be $E^c_0 |\tau_1-\tau_2|$ that gives the zero-order value of the Mott
   gap $\Delta_M = E^c_0.$ The bare interaction energy $E^c_0$ of two charges is reduced due
   to the presence of quadruples. In order to find this reduction it is convenient
   to consider the electric field created by the charge 1 and partially screened by the surrounding
   quadruples at the point $\tau_2$ of location of charge 2. In the limit of low
   quadruple density  the screening of the interaction due to different quadruples can be considered
   independently.

   It is easy to see that the electric field at the point $\tau_2$  is affected by the
   presence of the quadruple only if the later is placed such that charge $e_2$
   is inside the quadruple as it is shown in Fig.~\ref{quadruple_screening}. Depending on the arrangements
   of quadruple charges one can distinguish two different situations shown in Fig.~\ref{quadruple_screening}
   a) and b). In the case a) the electric field that charge $e_2$ feels is
   \begin{subequations}
   \label{field}
   \begin{equation}
   E^{(a)} = -3 E^c_0 + 2 E^c_{12} = -E^c_0 -
   E_{eh},
   \end{equation}
   while in the case b) it is
   \begin{equation}
   E^{(b)} =  E^c_0 - 2 E^c_{12} = -E^c_0 +E_{eh}.
   \end{equation}
   \end{subequations}
   We see from Eqs.~(\ref{field}) that the electric field at
   the point $\tau_2$ of location of charge 2 gets an additional contribution $\pm E_{eh}$ depending on
   the arrangements of quadruple charges. The probability of the configuration a)
   is given by the Gibbs weight as
   \begin{equation}
   \nonumber
   P_a(\tau_{q_1},\tau_{q_2}) =
   {\cal Z}_0^{-1}e^{-{\cal U}^q(\tau_{q_2}-\tau_{q_1})-E_{eh}(\tau_{q_2}-\tau_{q_1}) + \tau_{q_1}E_{eh}
   },
   \end{equation}
   where the time coordinates of  quadruple charges $\tau_{q_1}$ and
   $\tau_{q_2}$ are taken with respect to  coordinate $\tau_2$ where charge $e_2$
   is located such that $\tau_{q_2}>0$ and $\tau_{q_1}<0.$
   Analogously we find the probability of the configuration b)
   \begin{equation}
   \nonumber
   P_b(\tau_{q_1},\tau_{q_2}) = {\cal Z}_0^{-1}\, e^{-{\cal U}^q(\tau_{q_2}-\tau_{q_1})
     -E_{eh}(\tau_{q_2}-\tau_{q_1}) - \tau_{q_1}E_{eh} }.
   \end{equation}
   The probabilities $P_a$ and $P_b$ must be further multiplied by
   the quadruple density $\rho,$  coordination number $z$ and combinatoric factor 2.
   Multiplication by $\rho$ simply cancels the factor ${\cal Z}_0^{-1}$
   due to the relation $\rho={\cal Z}_0$ that we pointed out below
   Eq.(\ref{rho}). The additional contribution to the electric
   field at coordinate $\tau_2$ of charge 2 due to the presence of the
   quadruples becomes
   \begin{equation}
   \label{Delta_E}
   \Delta E = 2 z E_{eh} \int_{-\infty}^0 d\tau_{q1}  \int_0^{\infty}
   d\tau_{q_2} \; (P_b-P_a).
   \end{equation}
   The dielectric constant $\kappa_\tau$ is given by the expression
   $\kappa_\tau^{-1}=(E^c_0-\Delta E) / E^c_0.$ Taking integrals in Eq.~(\ref{Delta_E}) we finally get
   the dielectric constant
   \begin{equation}
\label{kappa_charge}
   \kappa_\tau^{-1} = 1 - {{ 2 g z }\over {\pi }}\, {{E_{eh}}\over {E_0^c }}\,
   \ln 2,
   \end{equation}
   that taking into account  relation~(\ref{delta_kappa})
   between the Mott gap and dielectric constant is indeed consistent
   with  result~(\ref{Delta_M_per_0}) previously obtained within the direct perturbation theory
   in tunneling matrix elements.

\section{Mott gap at large tunneling
conductances} \label{gap2}

In this section we continue our investigation of the  quadruple
model~(\ref{Charge1}) now coming to the regime of moderately large
tunneling conductance $g z \geq 1.$ We show that within this model
the Mott gap $\Delta_M$ is exponentially reduced with growth of
$g$ but nevertheless it always remains  finite and, thus, the
quantum system represented by the model~~(\ref{Charge1}) is always
an insulator at $T=0$ even at arbitrary large tunneling
conductance $g.$ This statement, however, turns out to be an
artifact of neglecting the high order tunnelling terms in
derivation of the phase action~(\ref{AES}) and the following from
it model~(\ref{Charge1}). Below we show that
models~(\ref{AES},\ref{Charge1}) are applicable as long as the
Mott gap $\Delta_M$ is larger than the inverse escape time of an
electron from a grain $\Gamma = g \delta.$ Thus, the results of
this section are applicable as long as the obtained Mott gap
$\Delta_M$ is larger than the inverse escape time $g \delta.$

\subsection{Mean field approximation}

The model~(\ref{Charge1}) can hardly be solved in  general case
and further in this section we make two assumptions that will
allow us to construct a mean field approach\cite{Herrero}. Namely,

   (i) we take the diagonal matrix of Coulomb
   interaction $E^c_{ij}=E^c_0 \, \delta_{ij}.$

   (ii) we assume that the coordination number $z$ is
   large.
\\
 These two assumptions allow to reduce the quadruple problem in
$d+1$ dimensions to
   the problem of one dimensional dipoles with self
   consistent dipole interaction.

   To construct a mean filed
   theory let us pick up a site $i$ and consider the influence of the
   neighboring sites on it in a mean field manner.
   The quadruple that occupies the  site $i$ and a neighboring
   site $j$ is reduced to the dipole located on the  site $i.$
   The mutual interaction of two charges of the resulting dipole
   consists of the quadruple interaction ${\cal U}^q$ and of
   the Coulomb interaction energy of quadruple charges located at
   the site $j.$  The later must be taken into account in a self-consistent
   way. The mutual influence of the neighboring sites due to
   direct Coulomb interaction is absent due to our choice of the
   diagonal Coulomb matrix $E^c_{ij}.$ Thus the internal dipole interaction
   energy becomes
   \begin{equation}
   \label{dipole_interaction}
   {\cal U}_d = \ln [ \pi ( \Delta\tau_n )^2 /g z] + \Omega_2(\Delta \tau_n),
   \end{equation}
   where  the interaction energy $\Omega_2(\tau)$ of two external
   charges must be calculated self consistently in
   terms of the dipole model. The asymptotic behavior of this function
   at the same time according to Eq.~(\ref{Omega_2}) determines the dielectric constant $\kappa_\tau$ that is directly
   related to the Mott gap $\Delta_M$ of the original model.

   Thus we arrive at the simplified one dimensional model of dipoles bounded by the internal
   interaction (\ref{dipole_interaction}). Apart from the bounding
   interaction all charges are subject to the
   1-d Coulomb interaction
   \begin{equation}
    U^c_{1d}={ E^c_0  \over 2} \sum_{n_1,n_2}
    |\tau_{n_1}-\tau_{n_2}| \, e_{n_1}\, e_{n_2}  \label{U}.
    \end{equation}

   In the limit of large density $\rho \gg 1$ that corresponds to the case
   of large tunneling conductance $g \gg 1$ the Coulomb interaction gets sufficiently screened.
   For this reason the expression for the density (\ref{rho}) as well as the
   identity ${\cal Z}_0 =\rho_0$ are applicable in the dense limit
   too.  In the dense limit
   the screening effects  can be usually well described within the Debay theory:
   \begin{figure}[tbp] \hspace{-0.5cm}
   \includegraphics[width=3.0in]{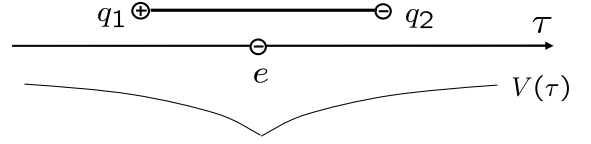}
   \caption{ Screening of the external charge $e$ in the dipole
    model. The Coulomb potential of the charge $e$ is screened by the
   dipole charges $q_1$ and $q_2$ resulting in the
   effective screened potential $V(\tau).$} \label{Debay}
   \end{figure}

   Consider the charge $e$ placed at $\tau=0$  shown in Fig.~\ref{Debay}.
   The Coulomb potential of this charge screened by the dipole charges
   obeys the Poisson equation
   \begin{equation}
   \label{Poison}
   \partial_\tau^2 V(\tau) = -2 E^c_0 \; ( q (\tau)-\delta(\tau)),
   \end{equation}
   where $ q(\tau)$ is the average density of all dipole charges
   and $\delta(\tau)$ is the delta function.
   At the same time the density of dipole
   charges is given by the Gibbs distribution: The contribution of
   two charges $q_1$ and $q_2$ of a single dipole
   ( located at coordinates $\tau_1$ and $\tau_2$ respectively )
   to the charge density $q(\tau)$ is
   \begin{equation}
   \label{prob}
   {\cal Z}_0^{-1}[\delta(\tau-\tau_1)-\delta(\tau-\tau_2)]\,
   e^{- \tilde {\cal U}_d(\tau_1-\tau_2)-V(\tau_1)+V(\tau_2) },
   \end{equation}
   where the energy $\tilde {\cal U}_d(\tau)$ that enters the Gibbs exponent includes the dipole
   interaction energy  ${\cal U}_d(\tau)$ and the screened Coulomb interaction of charges $q_1$
   and $q_2$
   \begin{equation}
   \label{35}
   \tilde {\cal U}_d(\tau) = {\cal U}_d(\tau)+\Omega_2(\tau).
   \end{equation}
   We note that within the approximation that we use the screened potential $V(\tau)$  is
   nothing but the effective screened interaction of two charges
   $\Omega_2(\tau)$ so that
   \begin{equation}
   \label{V_F_2}
   V(\tau) = \Omega_2(\tau).
   \end{equation}
   Multiplying Eq.~(\ref{prob}) by the dipole  density $\rho,$ making use of the identity
   ${\cal Z}_0=\rho$  and integrating over the coordinates $\tau_1,\tau_2$ we obtain the
   averaged charge density
   \begin{equation}
    \label{rho_tau}
   q(\tau) = {{2 g z}\over {\pi }}\, \int \limits_{-\infty}^{\infty} d\tau^\prime \,
   {{e^{ -2V(\tau-\tau^\prime)}}\over {(\tau-\tau^\prime )^2  }}
   \sinh [V(\tau^\prime) - V(\tau) ].
   \end{equation}

   Equations~(\ref{Poison}) and (\ref{rho_tau})  form a
   closed system of equations that determine the potential $V(\tau).$
   In the next subsection we present the solution of these equations.

\subsection{Solution of the mean field equations}

Numerical solution of Eqs.~(\ref{Poison},\ref{rho_tau}) for
moderately large tunneling conductances $g z \gtrsim 1$ shows that
the external charge is never screened completely and the long time
asymptotic behavior of the function
   $V(\tau)$ in Eq.~(\ref{V_F_2}) has a linear form
   \begin{equation}
   \label{asymp_F_2}
   V(\tau) =|\tau |  E^c_0 \, \kappa_\tau^{-1} \,   ,   \;\;\;\;\;  \tau \to
   \infty,
   \end{equation}
where the inverse dielectric constant $\kappa_\tau^{-1}$ becomes
very small for large conductances but nevertheless it remains
finite. The numerical solution also shows that the potential
$V(\tau)$ behaves logarithmically  on intermediate time scales and
then turns to the insulating linear behavior~(\ref{asymp_F_2}) on
longer times (crossover scale is given below).
  This suggests the following analytic approach to the solution of
  mean field Eqs.~(\ref{Poison},\ref{rho_tau}):
  To find the potential $V(\tau)$ on intermediate times where nonlinear effects are not
  important  we consider Eq.~(\ref{rho_tau}) in the linear approximation
  with respect to $V(\tau)$. As a result in the Fourier representation  we obtain
  \begin{equation}
  \label{rho_omega}
   q_\omega = - 2 z g |\omega | V_\omega.
  \end{equation}
  Together with the Poisson equation~(\ref{Poison}) written in the Fourier
  representation
  $
  -\omega^2 V(\omega) = - 2 {  E}^c_0 \,  (q_\omega -1)
  $
  equation (\ref{rho_omega}) defines the potential
  \begin{equation}
  \label{Vomega}
  V_{\omega} = - {1\over {\omega^2/2 { E}^c_0+ 2z g |\omega|
  }},
  \end{equation}
  that being transformed  to the time representation for  $\tau >(E^c_0 g )^{-1}$
  results in
  \begin{equation}
  \label{V_tau}
  V(\tau) = {{ \ln (\tau E^c_0 g) }\over { 2 \pi z g  }}.
  \end{equation}
The linear approximation that we used in Eq.~(\ref{rho_tau}) is
valid as long as the  potential $V(\tau)$ is small. This assumes
that dependence (\ref{V_tau}) is applicable for not very long
times $\tau \ll \exp(\pi z g) /g E_0^c.$  From the logarithmic
form of the answer (\ref{V_tau}) it follows that the change in the
charging energy
  \begin{equation}
  \label{scaling1}
  E_0^c \to \xi\, E_0^c
  \end{equation}
in  Eq.~(\ref{V_tau}) results in a constant shift of the potential
$V(\tau) \to V(\tau) + (1/2 \pi z g  ) \, \ln \xi. $ At the same
time the only term in Eq.~(\ref{rho_tau}) that is sensitive to
this shift is the potential $V(\tau)$ in the exponent that
provides  the renormalization of conductance
  \begin{equation}
  \label{scaling2}
  g\to g - (1/\pi z) \ln \xi.
  \end{equation}
  This consideration allows to solve the system of Eqs.~(\ref{Poison},\ref{rho_tau})
  for the case of large tunnelling conductance in the renormalization
  group(RG)  scheme. Since the charging energy $E_0^c$ is the only
  energy scale in the problem, the scaling of this energy with
  conductance determines the scaling of the Mott gap. Thus, from the
  scaling Eqs.~(\ref{scaling1},\ref{scaling2})
  we finally obtain  the expression for the Mott gap (\ref{DelM}).

  Equation~(\ref{DelM}) agrees with the result of RG approach of
  Ref.~[\onlinecite{Universal}] where the opening of the Mott gap
$\Delta_M$ was assumed to happen at the point where the tunneling
conductance running under RG scheme flowed to values of the order
one. We would like to note that the present approach does not rely
on any  assumptions of this kind: we first apply the RG scheme to
reduce the conductance to values $g z \geq 1,$ and further we
solve the equation numerically and show that opening of the gap
indeed takes the place, i.e. the asymptotics of the potential
$V(\tau)$ has a linear form indeed.

\section{Inclusion of higher order tunnelling terms}
\label{higher}

So far we considered the quadruple model (\ref{Charge1}) that
includes only the nearest neighbor electron tunnellings.
 In the present section we describe how the high order
tunneling terms can be included into our mapping scheme and in the
following sections we apply this technique to the problem of
hopping conductivity and Mott transition in granular arrays.

Keeping in mind the applications  to the irregular arrays, as in
the case of hopping conductivity, in this section we include the
random on-site  potential $\mu_i$ in the Hamiltonian into
consideration. Also we take into account variations of tunneling
conductance $g$ from grain to grain as well as variations of
charging energies $E^c_i$ and the mean energy level spacings
$\delta_i.$

In order to construct the effective classical model that takes
into account higher order tunnelling terms we proceed in the same
way as in the derivation of AES functional Eqs.~(\ref{AES}). First
we decouple the Coulomb interaction term in
Eq.~(\ref{hamiltonian}) with the help of potential field $V(\tau)$
  \begin{equation}
  \label{Coulomb_action} {\cal L}_c = - {1\over {2 e^2 }}\,
  \sum_{ij} [V_i + i\mu_i]  \, C_{ij} \, [V_j + i\mu_j] - i \sum_i
  n_i V_i,
  \end{equation}
  where  $n_i=n_i(\tau)$ is the electron density field. The
  term $-i n_i V_i$  can be absorbed  by the fermion gauge
  transformation given by Eqs.~(\ref{gauge}) such that the action
  that governs dynamics of the potential field $V(\tau)$ becomes
  \begin{equation}
  \label{Coulomb_action_new} {\cal L}_c = - {1\over {2 e^2 }}\,
  \sum_{ij} [V_i + i\mu_i]  \, C_{ij} \, [V_j + i\mu_j].
  \end{equation}
  \begin{figure}[t]
  \hspace{-0.5cm}
  \includegraphics[width=3.0in]{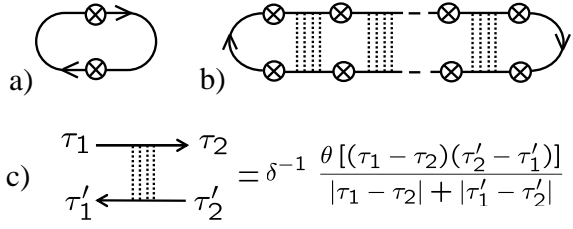}
  \caption{Diagram a) represents the lowest order correction in
  tunneling conductance $g$ to the partition function,
  the first term in the expansion (\ref{NN_Partition_Function}).
  The solid lines denote the propagator of electrons,
  the crossed circles stand for the phase dependent tunnelling
  matrix elements $t_{ij}e^{i\phi_{ij}(\tau)}.$ Diagram b) describes
  a higher order correction in tunnelling matrix elements $t_{ij}.$ The dotted
  lines represent impurity scattering. The diagram c) shows the zero
  dimensional diffusion in time representation.
   } \label{Diagrams}
  \end{figure}
Further, following Ref.~[\onlinecite{Beloborodov01}] we proceed
with diagrammatic expansion of the partition function $Z$ in
tunnelling matrix elements  $ t_{ij} \,e^{i\phi_{ij}(\tau)}$. In
the lowest order we reproduce the results obtained within the
phase functional. Indeed, the lowest order correction to the
partition function $Z$ shown in Fig.~\ref{Diagrams}a simply
results in the first term in the expansion
(\ref{NN_Partition_Function}). All disconnected diagrams of the
type shown in Fig.~\ref{Diagrams}a must also be included in the
partition function,  this reproduces all high order terms in the
expansion~(\ref{NN_Partition_Function}). Integration over the
phase fields $\phi_i(\tau)$ can be implemented exactly as it was
described in Sec.~\ref{b} resulting in the appearance of classical
charges in $d+1$ dimensions with the Coulomb interaction energy
\begin{equation}
U={1\over 2} \sum_{n_1,n_2} E^c_{i_{n_1} i_{n_2}}
|\tau_{n_1}-\tau_{n_2}| \, e_{n_1}\, e_{n_2} +\sum_n \mu_{i_n}
\tau_n e_n  \label{U_mu}.
\end{equation}
that differs from Eq.~(\ref{interaction}) only by the presence of
the random potential $\mu_i$ not previously taken into account.
\begin{figure}[tbp]
\hspace{-0.5cm}
\includegraphics[width=3.0in]{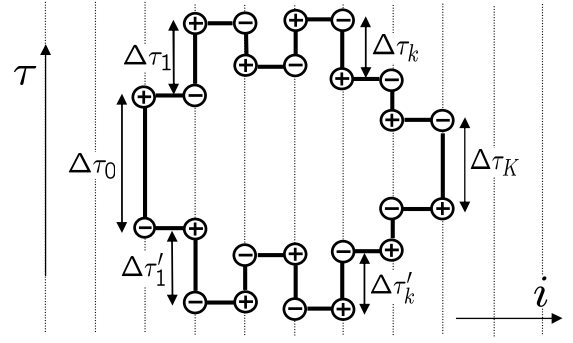}
\caption{Illustration of the mapping of the model of quantum
electrons in a granular array onto the model of classical charges
imbedded into the electron world lines forming loops. The
illustration is shown for the case of $1+1$ dimensions ($1d$
granular array),  generalization to higher $d+1$ dimensions is
straightforward.} \label{Loop}
\end{figure}
Thus, inclusion of all disconnected diagrams,
Fig.~\ref{Diagrams}a, and their further averaging over the phases
reproduces the classical quadruple model~(\ref{Charge1}). The
expression for the internal quadruple energy for the case where
conductance $g$ varies from contact to contact is the
straightforward generalization of Eq.~(\ref{qudruple_energy})
\begin{equation}
 \label{internal_energy_1}
 {\cal U}^q_n = \ln [ 2 \pi ( \Delta\tau_n )^2 /g_{ij} ].
 \end{equation}

The advantage of the mapping of the original quantum model on the
classical electrostatic system is that it allows to include the
higher order tunneling processes shown in Fig.~\ref{Diagrams}b in
essentially the same way as the nearest neighbor hoppings. The
higher order diagrams include the single grain diffusion
propagator
\begin{equation}
D_0 = 2\pi \delta^{-1} /|\omega_1 - \omega_2|,
\end{equation}
 that being transformed into the time representation results in the expression
shown in Fig.~\ref{Diagrams}c. Each tunneling matrix element in
the diagram Fig.~\ref{Diagrams}b  includes the phase variables as
$t_{ij}e^{i\phi_{ij}(\tau)}.$ This allows to present the diagram
in Fig.~\ref{Diagrams}b as a charge loop shown in Fig.~\ref{Loop}.
The charges interact electrostatically and are subject to the
local potentials according to Eq.~(\ref{U_mu}). The internal
energy of a single $K-$th order loop is
\begin{eqnarray}
\label{K_loop}
 E^{(K)} & = &  \ln [ 2\pi K/g_0] -
\sum_{k=1}^{K-1} \ln [ g_k \delta/ 2\pi ] + \ln
|\Delta\tau_0|  \nonumber \\
&& + \ln|\Delta \tau_K| + \sum_{k=1}^{K-1} \ln ( |\Delta\tau_k| +
|\Delta\tau_k^\prime| ),
\end{eqnarray}
where the time intervals $\Delta\tau_k$ and $\Delta\tau^\prime_k$
shown in Fig.~\ref{Loop} are restricted to have opposite signs,
i.e. $\Delta \tau_k \Delta\tau^\prime_k <0.$

We mapped the model of quantum electrons in a metallic array with
Coulomb interaction onto the classical model of Coulomb charges
bounded  by the electron world lines and forming loops. First
order loops that form quadruples represent the process of virtual
electron tunnelings to the neighboring grains and back. Higher
order loops represent electron motion on larger distances.

In the Mott insulator state electrons cannot move on large
distances. This assumes that in the classical representation of
the Mott insulator state the thermodynamics of the classical
charges will be dominated by the contribution of the short loops,
while high order loops will be energetically suppressed.

In the metallic state, on the contrary, diffusive motion of
electrons on any distances is possible meaning that in the
classical model the loops of infinite order are present. It is
clear that the insulator to metal transition in terms of the
classical model corresponds to the transition at which the
infinite loops that were suppressed in the insulating state begin
to proliferate. Infinite loops represent almost free charges that
screen the Coulomb interaction.  The inverse dielectric constant
$\kappa_\tau^{-1}$ of this state is zero that corresponds to the
absences of the Mott gap $\Delta_M$.

The detailed description of the Mott transition is difficult since
it requires the analysis of contribution of high order loops in
the regime where the tunneling conductance $g$ is not small and
the Mott gap $\Delta_M$ is already strongly reduced due  to the
quadruple screening according to Eq.~(\ref{DelM}). Such
description, on the qualitative level is performed in
Sec.~\ref{Mott}, while in the next section we will study the high
order electron tunnellings in a more simple regime of low
tunneling conductance. As we discussed above, the high order
tunnelings being not important for the thermodynamics in the
regime of small $g$ are crucial for the description of the hopping
transport in the case of irregular array.

\section{Hopping conductivity in Granular metals}
\label{disorder}

In Sec.~\ref{higher} it was shown that the periodic granular array
with low tunneling conductances $g$ is the Mott insulator with the
hard gap $\Delta_M$ in the electron excitation spectrum. This
assumes the activation conductivity behavior, $\sigma \sim
\exp{(-\Delta_M/T})$, and excludes the possibility of hopping
transport. However, as we pointed out in the Introduction, in
realistic granular arrays there are many reasons for electrostatic
disorder that compensates  the Coulomb blockade on some sites.
This gives rise to the finite density of states on the Fermi level
and opens the possibility for the variable range hopping
transport.

In this section we discuss the hopping conductivity in granular
metals via elastic and inelastic electron cotunneling. We first
concentrate on hopping via elastic cotunneling and later discuss
the inelastic mechanism.

\begin{figure}[tbp]
\hspace{-0.5cm}
\includegraphics[width=3.5in]{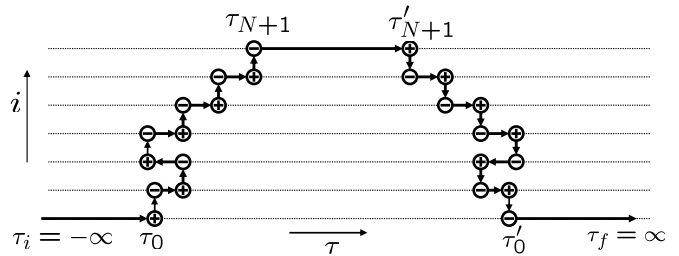}
\caption{ The electron world line representing the probability of
elastic cotunneling from the $0-$th to the $N+1-$st grain. }
\label{tunneling}
\end{figure}

\subsection{Hopping via elastic cotunneling}
\label{elastic_hopping}

To analyze the hopping transport mechanism in granular array one
has to consider the probability of electron tunnelling between the
two distant grains with energies close to the Fermi level. At very
low temperatures such a tunneling can be realized only as a
virtual tunneling of an electron through several grains being in
the Coulomb blockade regime - such a process is called the elastic
cotunneling.~\cite{Averin}

In the following we consider the probability of elastic
cotunnelling through several grains within the approach of
effective classical model. We assume that the electron tunneling
takes place between two sites $i_0$ and $i_{N+1}$ where the local
Coulomb gap is absent. In our model it means that the Coulomb
energy on these two sites is compensated by the external local
potentials $\mu_i.$ Thus, removing the charge from the grain $i_0$
as well as placing it on the grain $i_{N+1}$ does not cost any
Coulomb energy. The probability of tunnelling between these two
states can be found as follows: In the basis of the eigenstates of
a system of isolated grains the amplitude of tunnelling process
from the site $i_0$ to the site $i_{N+1}$ is
 \begin{equation}
 A_{i_0,i_{N+1}} = <0| \; \hat c_{i_{N+1}}  \hat S  \; \hat c^\dagger_{i_0}  \,
 |0>,
 \end{equation}
 where $\hat S$ is the evolution operator that written in the interaction representation
 takes into account only tunnelling part of the Hamiltonian~(\ref{hamiltonian}).
 Consideration of the amplitude  $A_{i_0,i_{N+1}}$, however, is not convenient
 since the amplitude becomes zero after averaging over tunnelling matrix elements $t_{k_1,k_2}.$
 This happens because the phase of the amplitude of tunnelling  process is very sensitive
 to the particular realization of disorder.  For this reason one has to consider the probability
 of the tunnelling process $P=A^* A.$ The
 conjugate amplitude $A^*$ can be written as the probability of the
 inverse process of tunnelling between $i_{N+1}-$st and $i_0-$th grains
 \begin{equation}
 A_{i_0,i_{N+1}}^* = A_{i_{N+1},i_0} =  <0| \; \hat c_{i_0} \, \hat S  \; \hat c^\dagger_{i_{N+1}}  \,
 |0>.
 \end{equation}

 The probability of the tunnelling process
 $P=A^* A$ can be presented as an electron world-line shown in
 Fig. \ref{tunneling}. The first interval of the world line ($\tau_i,
 \tau_0$) represents the initial state of the system where the
 electron is located at the site $i_0.$ The
 following part ( $\tau_0, \tau_{N+1} $ ) represents
 the amplitude of the tunnelling process $ A_{i_0,i_{N+1}} $ from the
 site $i_0$ to the site $i_{N+1}.$ The interval ($\tau_{N+1}, \tau_{N+1}^\prime $)
 describes the system with the electron located on the
 site $i_{N+1}.$ The next   interval ($ \tau_{N+1}^\prime, \tau_0^\prime $) represents
 the amplitude  $A_{i_{N+1},i_0}$ of the process of tunnelling back to the site $i_0,$ and the last
 interval  ($\tau_0^\prime, \tau_f$) represents the system back in the initial state where the
 electron is located at site $i_0.$

The world line that represents the tunnelling probability $P =
A^*A$ is very similar to the closed loops in Fig.~\ref{Loop}. The
only difference is that in the world line of Fig.~\ref{tunneling}
the Green functions on the sites $i_0$ and $i_{N+1}$ must be taken
in a different way: In order to capture the energy resolution of
electron energies of the initial and final states we will assume
that both sites $i_0$ and $i_{N+1}$ have only one quantum state
each with energies $\xi_0$ and $\xi_{N+1}$ (counted with respect
to the chemical potential) respectively.

When electron tunnels from site $i_0$ to  site $i_1,$ the hole is
created at site $i_0.$ Such process is described by the Green
function of a negative time argument \cite{Abrikos}
  \begin{equation}
  G_0 (\tau_{0}^\prime-\tau_{0}) = n(\xi_0) \; e^{
  \xi_{0}(\tau_{0}^\prime-\tau_{0})},
  \end{equation}
  where  $n(\xi)$ is the electron occupation number that reminds that the electron is
  assumed to be initially present at site $i_0$.

  Similarly, the process when the electron is located on  site
  $i_{N+1}$ is described by the Green function of positive
  arguments
  \begin{equation}
  G_{N+1} (\tau_{N+1}^\prime-\tau_{N+1}) =   [\,1- n(\xi_{N+1})\, ]\; e^{-\xi_{N+1}
  (\tau_{N+1}^\prime-\tau_{N+1}),
  },
  \end{equation}
where the occupation number $1- n(\xi_{N+1})$ insures that the
final state must be empty for an electron to be able to tunnel
there.

Before proceeding to the calculation of tunneling probability let
us first fix notations for all time coordinates in
Fig.~\ref{tunneling}: To distinguish the time coordinates that
belong to the left part of the world line representing  the direct
tunneling process  from the ones that belong to the right part
representing the inverse process we denote the later ones by
primes. Further, to distinguish the time coordinates of positive
and negative charges we use the index $+$ and $-$ respectively.

The tunneling probability $P$ represented by the world-line shown
in Fig.~\ref{tunneling} can be calculated easily for the case of a
diagonal (short range) Coulomb potential $E^{\scriptscriptstyle
C}_{ij} = \delta_{ij} E^c_i.$

\subsubsection{Short range Coulomb interaction} \label{elastic}

In this case time integrations on each site $k$ can be implemented
independently and the tunneling probability $P$ is given by the
product of probabilities $P_k$ representing the single site
contributions. To find a single site contribution $P_k$ we note
that there are two possible arrangements of the tunneling times:
(i) $\tau_{k-} < \tau_{k+} < \tau_{k-}^\prime < \tau_{k+}^\prime $
and (ii) $\tau_{k+} < \tau_{k-} < \tau_{k+}^\prime <
\tau_{k-}^\prime .$ Let us first consider the case (i): The
Coulomb energy of such configuration is given by
 \begin{equation}
 \label{Uk}
 U_k^{c(i)} = E_k^+ \,( \Delta\tau_k +
 \Delta\tau_k^\prime ),
 \end{equation}
where the electron excitation energy $E_k^+$ is given by
Eq.(\ref{eh}) and the time intervals
$\Delta\tau_k,\Delta\tau_k^\prime
> 0$ are defined as $ \Delta\tau_k = \tau_{k+} - \tau_{k-}, \;\;
\Delta\tau_k^\prime = \tau_{k+}^\prime - \tau_{k-}^\prime. $  The
internal energy of the configuration according to
Eq.~(\ref{K_loop}) is
 \begin{equation}
 \label{eps_k}
 {\cal U}_k = - \ln[g_k \delta /2\pi ] +\ln(\Delta \tau_k +\Delta\tau_k
 ^\prime),
  \end{equation}
 where $g_k$ is the tunnelling conductance between the $k-$th and
 $(k + 1)-$st grains. One can see that the second arrangement of charges  (ii)
 differs only by the expression for the Coulomb energy
\begin{equation}
 \label{Uk_}
 U_k^{c(ii)} =  E^-_k \, ( |\Delta\tau_k| +
 |\Delta\tau_k^\prime| ).
 \end{equation}
Now with the help of Eqs.~(\ref{Uk}, \ref{eps_k}, \ref{Uk_}) we
find the elementary probability $P_k$ determined by the sum of the
contributions (i) and (ii) as
 $$
 \label{pk} P_k = {{ g_k \delta }\over{ 2 \pi }}
 \int_{0}^\infty d\Delta\tau\,  d\Delta\tau^\prime\,  {{ e^{-E^+_k(\Delta \tau +\Delta\tau^\prime)} +
 e^{-E^-_k( \Delta \tau+\Delta\tau^\prime)  }} \over{\Delta\tau + \Delta\tau^\prime
 }},
 $$
and taking the integrals  we obtain
\begin{equation}
 P_k = {{g_k \, \delta}\over {  \pi   \tilde E_k } } .
 \end{equation}
where the energy $\tilde E_k$ is the combination of the energies
$E^+_k$ and $E^-_k$ defined by Eq.(\ref{tilde_E}).

 To find the time dependence of the tunneling process we notice that the
time intervals $\tau_{0}^\prime-\tau_{0}$ and
$\tau_{N+1}^\prime-\tau_{N+1}$ coincide within the accuracy of the
inverse charging energy.  In the limit that we consider the
charging energy is the largest scale in the problem that allows to
take both intervals to be equal to the the "width" of the loop
$\tau_w$ that at the same time represents the time that electron
spend on site $i_{N+1}.$ Thus, the time dependence of the
tunneling process is simply given by
 \begin{equation}
 \label{time_dependence}
 e^{-(\xi_{N+1}-\xi_0)\tau_w}.
 \end{equation}
Making an analytical continuation to real times $\tau_w = i t_w$
and taking the last integral over $t_w$ we obtain the delta
function
\begin{equation}
2\pi \delta(\xi_{N+1}-\xi_0 ),
\end{equation}
that indeed shows that the process is elastic, i.e. electron can
tunnel only to the state with exactly the same energy.

 Finally, for the tunneling probability of the elastic process we
 obtain
\begin{equation}
 \label{P} P_{el} = w\, \delta(\xi_{N+1}-\xi_0  ) \,g_0 \,  \prod_{k=1}^{N}
 P_k,
 \end{equation}
where the factor $ w = n(\xi_0)(1-n(\xi_{N+1})$ takes into account
   the occupation numbers of the initial and final states.
 We see that the total probability $P_{el}$ contains the product of all
conductances, mean level spacings and inverse energies along the
tunneling path. For this reason it is convenient  to introduce the
geometrical averages of theses quantities along the tunneling path
that were introduced in Sec.~\ref{sum_el_cot}. Writing the total
probability $P_{el}$  in terms of the quantities $\bar g,
\bar\delta, \bar E$ we obtain the result (\ref{P_el_result}).

\subsubsection{Long range Coulomb interaction}

In the presence of the long range Coulomb interaction the
situation is more complicated since  the integrals over variables
$\tau_{\pm k}$ cannot be taken on each site $k$ independently.
However, one can estimate the tunnelling probability $P_{el}$ by
finding its upper and lower limits: Indeed, neglecting the long
range part of the potential (as we did above) one gets the upper
boundary. The lower boundary can be obtained by considering such
trajectories where all $\Delta \tau_k$ are of the same sign. In
this case the electric field created by the charges on step $k$
does not interfere with other charges and thus one can implement
integrations independently. This gives a lower boundary for the
tunnelling probability $P_k$ which is about a factor 2 smaller
than the tunnelling probability $P_k$ obtained neglecting the long
range part of the Coulomb potential. This results in the
boundaries of the factor $0.5 \lesssim c < 1.0$ in the expression
for localization length $\xi_{el}$ in Eq.~(\ref{localization}).

The above derivation of tunnelling probability $P_{el}$ has been
done for the zero temperatures case.  Of course the finite
temperatures and the presence of phonons are necessary for the
realization of the Mott-Efros-Shklovskii mechanism. Calculation of
the tunneling probability  at zero temperature is justified as
long as inelastic  tunneling   processes that are considered in
the next section  can be neglected.

\subsection{Hopping via inelastic cotunneling}

According to the theory of transport through a single quantum dot
in the Coulomb blockade regime~\cite{Averin}, at temperatures $T>
\sqrt{\delta E_C}$ the electron transport is dominated by
inelastic cotunneling processes. In this case the process of
charge transfer from one contact to the dot and the following
process of charge transfer from the dot to the other contact is
realized by means of different electrons. As a result, the energy
of the outcoming electron differs from that of the incoming one
and the dot energy is changed after the each elementary charge
transfer process. For this reason  inelastic cotunneling processes
are allowed only either at  finite temperature or if a finite
voltage is applied to  contacts.

 \begin{figure}[tbp]
 \hspace{-0.5cm}
 \includegraphics[width=3.5in]{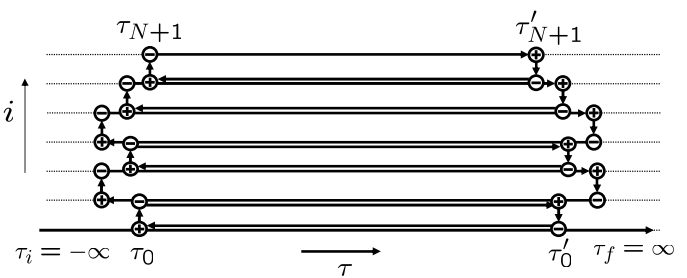} \caption{ Tunneling
 via inelastic cotunneling between  $0-$th to the $N+1-$st grains. }
 \label{inelastic}
 \end{figure}

In order to include the inelastic cotunneling processes into the
scheme that we described above, first,  we  have to generalize the
charge representation to finite temperatures. The main difference
is that the imaginary time $\tau$ at finite temperatures is
constraint to the interval ($0, \beta),$ where $\beta=1/T$ is the
inverse temperature. Also the expressions for internal energies
must be properly modified. Using the Matsubara technique one can
show that the proper expressions for the internal energies can be
obtained from the corresponding expressions at zero temperature
via the substitution $ 1 / \tau \to \pi T /  \sin( \pi T \tau ), $
such that, for example, the quadruple internal energy given by
Eq.~(\ref{qudruple_energy}) becomes
  \begin{equation}
  \label{E}
  {\cal U}_n^q = \ln [ \, 2 \sin^2( \pi T | \Delta \tau |)/ \, g \pi T^2\,
  ].
  \end{equation}
In general,  one should also take into account the static charges
that formally appear from the integer winding numbers in the phase
representation at finite temperature. However, inclusion of static
charges is not needed for the description of virtual co-tunneling
processes that we consider.

The representation of an inelastic cotunneling process in terms of
the electron world lines is shown in Fig.~\ref{inelastic}. This
process resembles the world line of the elastic co-tunneling
process shown in Fig.~\ref{tunneling} since the electron charge is
transferred from the site $i_0$ to the site $i_{N+1}$ and then
back to the site $i_0$ in essentially the same way. However, the
important difference is that the diagram for the inelastic process
is constructed out of quadruple loops only. This implies that the
charge is transferred by means of deferent electrons on each hop.

As in the case of the elastic cotunneling we  assume that the
sites $i_0$ and $i_{N+1}$ have only one state each with energies
$\xi_0$ and $\xi_{N+1}$ respectively. Similarly to the previous
subsection consideration, we write down the Green functions
representing the initial and final states: when electron tunnels
from site $i_0$ to site $i_1,$ a hole is created at site $i_0$.
Such a process is described by the Green function of negative time
arguments
  \begin{equation}
  G_0 (\tau_{0}^\prime-\tau_{0}) = n(\xi_0) \; e^{
  \xi_{N+1}(\tau_{0}^\prime-\tau_{0})}.
  \end{equation}
The process where electron is  located on the site $i_{N+1}$ is
described by the Green function of positive arguments
   \begin{equation}
   G_{N+1} (\tau_{N+1}^\prime-\tau_{N+1}) =   [\,1- n(\xi_{N+1})\, ]\; e^{-\xi_{N+1}
   (\tau_{N+1}^\prime-\tau_{N+1}) }.
   \end{equation}

The time coordinates $\tau_{+k}$ and $ \tau_{-k} $ are again
confined within the temporal interval $\Delta \tau_k\sim 1/
E^c_0$.  This means that the time arguments of the quadruple
energies ${\cal U}_n^q$ in Eq.~(\ref{E}) are approximately the
same and equal to the ``width" of the loop formed by the world
line in Fig.~\ref{inelastic}: $\tau_w\approx\tau_{0}^\prime -
\tau_0 \approx \tau_{N+1}^\prime - \tau_{N+1}$. Thus, the
contributions of quadruple internal energies ${\cal U}_n^q$ to the
tunneling probability $P_{in}$ result in the factor
   \begin{equation}
   \left( {{ \bar g \pi T^2 }\over {\sin^2(\pi T \tau_w )  }}   \right)^N,
   \end{equation}
where $\bar g$ is the geometrical average of the conductances over
the electron path. Integrations over the intermediate times
$\tau_k$ of the inelastic process can be easily done for the case
of the diagonal Coulomb interaction $E_{ij}^c = E_0^c\delta_{ij}$.

\subsubsection{Short range Coulomb interaction} \label{inelastic_sr}

In the case of the short range interaction the integrations over
$\Delta \tau_k$ and $\Delta \tau^\prime_k $ can be performed
independently.  Also the integrations over the $\Delta\tau_k$ and
$\Delta\tau^\prime_k$ result in the equal contributions given by
   \begin{equation}
   \label{Coulomb_in}
   \int_{-\infty}^{\infty}d \Delta\tau_k \; e ^{ - E_k^c |\Delta\tau_k| +\mu_k \Delta\tau_k}=
   2/\tilde E_k  ,
   \end{equation}
where the energy $\tilde E_k$ is defined in Eq.~(\ref{tilde_E}).
Collecting all the terms for the probability $P_{in}$ one obtains
   \begin{equation}
   \label{analit}
   P_{in}(\tau_w) = {{   w \, \bar g^{N+1}  } \over {2 \pi^{N+1} }}
   \left( {{ 2 \pi T }\over { {\bar E} \sin(\pi T \tau_w) }} \right)^{2N}
   e^{-\Delta \tau_w }.
   \end{equation}
Finally, in order to find the tunneling probability $P_{in}$ via
the inelastic cotunneling one has to make the analytical
continuation in Eq.~(\ref{analit}) to the real times $\tau_w = i
t_w$ and integrate over $t_w$ arriving at
   \begin{equation}
   \label{P_in}
   P_{in} = {{   w \, g^{N+1}  } \over {2 \pi^{N+1} }} \int \limits_{-\infty}^{+\infty} dt
   \left[ {{ 2 \pi i T }\over {  {\bar E} \sinh(\pi T t )}}
   \right]^{2N} e^{-i \Delta t }.
   \end{equation}
Here the singularity of the function $\sinh^{-2N}(\pi T t)$ is
taken in the upper half of the complex plain. At zero temperature
one can easily calculate the integral over the variable $t$ in
Eq.~(\ref{P_in}) by shifting the contour of integration in the
complex plain either to negative complex infinity  in the case of
positive $\Delta$ or to the positive complex infinity  in the case
of negative $\Delta.$ In the first case we obtain $P_{in}=0$. This
reflects the fact that the real tunneling process with increase of
electron's energy is forbidden at $T=0$. In the latter case,
$\Delta < 0$, the zero temperature probability is determined by
the pole of the function $\sinh^{-2N}(\pi T t)  $ that results in
the answer presented in Eq.(\ref{inel}).

At finite temperatures the integral in Eq.~(\ref{P_in}) can be
expressed in terms of the  the Euler Gamma functions that leads to
the general expression (\ref{result_in1}).

\subsubsection{ Long range Coulomb interaction} \label{inelastic_lr}

In the presence of the long range part of the Coulomb interaction
the integrations over time variables on each step cannot be
performed independently. This makes it difficult to derive an
exact analytic formula for the tunneling probability. However, as
in the case of the elastic cotunneling, one can easily estimate
the upper and lower boundaries for this quantity: neglecting the
off-diagonal part of the Coulomb interaction, as above, we get the
upper boundary. The lower boundary can be obtained by considering
the trajectories where all $\Delta\tau_k$ (as well as all
$\Delta\tau_k^\prime$) have the same sign. This consideration
constraints the values of the coefficient $c$ in the expression
for the localization length to the interval $1/4 \lesssim  c<1$.
One can also see that only the sites that are geometrically close
to each other can interfere. Indeed, let us consider two sites
$k_1$ and $k_2$ in the tunneling path that are relatively far from
each other. Interference effects related to these sites can be
included as a small correction. One can easily see that the first
order correction vanishes because the average electric field on
the site $k_1$ created by the charges on the site $k_2$ is zero
and vias versa. Thus the contribution of the interference effects
falls of with the distance as $1/r^2$ at least. These insures that
there is no any dramatic effect related to the long-range part of
the Coulomb interaction apart form the numerical correction in the
localization length. The same arguments apply to the case of the
elastic cotunneling.

\section{ Mott transition in a periodic granular array } \label{Mott}

\label{Mott_transition}

In this section we discuss the metal to insulator transition in
the regular periodic granular arrays.  The Mott gap $\Delta_M$ for
such a system in the nearest neighbor hopping (quadruple)
approximation considered in Sec.~\ref{gap2} is exponentially small
as function of conductance, Eq.~(\ref{DelM}), but always finite.
In this section we consider how this result is affected by
inclusion of high order tunneling processes that are represented
by higher order loops in the classical model of Coulomb charges.

In Sec.~\ref{disorder} it was shown that at $g z \ll 1$ the
contribution of higher order tunnellings processes to the
tunneling provability $P$ at zero temperature is suppressed by the
factor
\begin{equation}
\label{small_factor}
 \left( \frac{g \, \delta }{E^c_0} \right)^N,
\end{equation}
where  $N$ is the order of the tunneling process. It is clear that
the same parameter defines the smallness of the contribution of
the high order loops to the thermodynamic potential $\Omega$ of
the classical gas. If the tunneling conductance is not small, $g z
> 1$, the Coulomb interaction of two classical charges is strongly
renormalized and in this case instead of the bare Coulomb
interaction $E_0^c$ in Eq.~(\ref{small_factor}) one has to take
the renormalized one - that is the Mott gap $\Delta_M$ given by
Eq.~(\ref{DelM}). We come to the conclusion that the high order
loops can be neglected as long as
\begin{equation}
\label{Gap>gd} \Delta_M > g \delta.
\end{equation}
Equation~(\ref{Gap>gd}) can be rewritten in terms of the
intergranular tunneling conductance $g$ in the way
\begin{equation}
g < g_c,
\end{equation}
where the critical conductance $g_c$ is defined by
Eq.(\ref{critical_conductance}).  The samples characterized by the
tunneling conductance $g < g_c$ at zero temperatures are
insulators with the Mott gap $\Delta_M$ given by Eq.~(\ref{DelM}).
The result~(\ref{critical_conductance}) for the critical
conductance $g_c$ coincides with the value of the stability of the
metallic state obtained from the diagrammatic technique from the
metallic side of the transition $g > g_c.$~\cite{BLV03} This fact
confirms that the value (\ref{critical_conductance}) indeed
separates the metallic and insulating states at $T=0.$

Note that the accuracy of the above considerations is not
sufficient to establish whether the transition between the
insulating and metallic states is continuous or discontinuous. The
details of the behavior of the Mott gap $\Delta_M$ in the
immediate vicinity of the critical conductance $g_c$ are not known
yet.

So far we considered the limit of the zero temperature. Physics of
the Mott transition at finite temperature depends crucially on
whether the transition is continuous at $T=0$ or it is
discontinuous.  In the case of continues transition at $T=0$, at
finite temperatures the metallic and insulating phases, strictly
speaking, cannot be distinguished. Indeed, the Mott transition
does not break any symmetry and the conductivity at finite
temperatures is finite even in the "insualting" phase where it
acquires the activation form, Eq.~(\ref{activation}). However, in
the case of the first order phase transition at $T=0$, due to
arguments of physical continuity, the transition has to remain of
the first order in some interval of finite temperatures. Such
transition would result in the jump of the physical observables
even at finite temperatures.

\section{Conclusions}
\label{conclusions}

We have investigated Coulomb interaction effects and transport
properties of granular conductors in the insulating (dielectric)
regime. We have considered both the periodic- and irregular arrays
and took into account the  random on-site electrostatic potential.
We find the dependence of the Mott gap on the tunneling
conductance in the dielectric regime of a regular array. The
insulator to metal transition takes place at the critical
conductance given by Eq.(\ref{critical_conductance}) where the
Mott gap becomes of the order of the inverse escape time of the
electron from a single grain $g\delta$. The order of the
transition remains, however, an open question. For the case of
irregular granular arrays we have derived the Efros-Shklovskii law
for the conductivity of a granular system. Depending on the
temperature the hopping conductivity is dominated by either
elastic or inelastic cotunneling processes. In the latter case the
localization length is weakly (logarithmically) depends on
temperature.  In particular, our model and the obtained results
explain the origin of the ES conductivity in the spatially
periodic arrays of semiconducting quantum dots.

Finally, we would like to discuss the possibility of the
observation of the Mott law in granular metals. To remind, in
semiconductors, the Efros-Shklovskii law may turn to the Mott
behavior with the increase of temperature. This happens when the
typical electron energy  $\varepsilon$  involved  in a hopping
process becomes larger than the width of the Coulomb gap
$\Delta_c$, i.e. when it falls into the flat region of the density
of states where Mott behavior is expected. To estimate the width
of the Coulomb gap $\Delta_c$ - one compares the ES expression
(\ref{ES_density_of_states}) for the density of states with the
bare density of states $\nu_{g_0}$, i.e. the DOS in the absences
of the long-range part of the Coulomb interactions obtaining
\begin{equation}
\label{crossover}
 \Delta_c = \left( \frac{\nu_{g_0} e^{2d}}{\tilde \kappa^d} \right) ^{1\over
{d-1}  } .
\end{equation}
As we discussed in Sec.~\ref{summary} in granular metals the
density of states $\nu_{g_0}$ that is relevant for hopping
conductivity does not coincide with the physical DOS, since only
the lowest energy excitations in a single grain must be counted in
determination of $\nu_{g_0}.$~\cite{Shklovskii04}

Bare density of ground sates  $\nu_{g_0}$  can be easily estimated
for the case of strong on-site disorder that we assumed so far as
follows:  The width of the potential distribution $\mu_i$ is
approximately given by the average on-site Coulomb energy so that
\begin{equation}
\nu_{g_0} \approx 1/ E^c_0 \, a^d.
\end{equation}
Inserting this value into Eq.~(\ref{crossover})  for the width of
the Coulomb gap we obtain $ \Delta_c \sim E^c_0  $ meaning that
there is no flat region in the density of ground states and, thus,
the Mott regime is not possible.  However, the Mott regime may
appear in the case of weak electrostatic disorder ( for a detail
discussion of this issue see Ref.~\cite{Shklovskii04}).  In this
case the subgap states will appear very rarely over the granular
sample and the bare density of ground states $\nu_{g_0}$ is
greatly reduced and as a result the width of the Coulomb gap
substantially shrinks.

One can easily repeat all the considerations that led us to the
expressions for the hopping conductivity for the case of flat
density of ground states arriving to the Mott expression
\begin{equation}
\label{Mott_law}
 \sigma \sim  \sigma_0 \; \exp [-(T_{0}^M/T)^{1\over {d+1}}],
\end{equation}
where the temperature $T_0^M$ is given by
\begin{equation}
T_0^M \sim 1/  \nu_{g_0} a^d  \xi^d,
\end{equation}
with the dimensionless localization length $\xi$ given by either
equation (\ref{localization}) or (\ref{inelastic_loc_len})
depending on the temperature regime.

The crossover temperature $T^*$   between the two regimes is
easily obtained exactly as in the case of semiconductors; by
comparing Eqs.~(\ref{hopping}) and (\ref{Mott_law}) and using the
corresponding expressions for $T_{\circ}$ we get
\begin{equation}
T^*\sim
e^2 a \xi \left( {{ e^4 \nu_{g_0}^2 }\over {\tilde \kappa ^{d+1}
}} \right)^{1/(d-1)}.
\end{equation}

 As a final note, we would like to comment on the possibility of
 fixing the numerical coefficient in the
 expression for the temperature $T_0$ that enters exponents of the
 expressions for the hopping conductivity. In the case of hopping conductivity
 in semiconductors this coefficient can be fixed by means of the mapping the problem of the impurity
 states in semiconductors onto a simple classical percolation model that can be investigated completely
 numerically\cite{Abrahams,Halperin,Shklovskii}. In granular metals, on the qualitative level, the situation is
 very similar, however, it is not clear that a model of a granular metal can be reduced to a simple unique
 percolation problem without making additional uncontrolled assumptions. The main complications are that the
 tunneling process in granular metals (i) in general goes via the curved trajectory, (ii) it depends on the
  number of
 the parameters of the intermediate grains in the tunneling path as $ E^c_i, \delta_i,
 g_{ij}.$ For this reason, in general, we expect that the coefficient in the expressions for
 $T_0$ is not universal and is dependent on the morphology of a granular sample.

After the present study was completed, we became aware of the
preprint \cite{FI05} where the hopping transport in granular
metals via elastic and inelastic cotunneling  in the linear regime
( low applied voltage )  was considered; their results essentially
agree with ours.

\begin{acknowledgments}
We acknowledge useful discussions with  A.~E.~Koshelev,
V.~I.~Kozub, K.~A.~Matveev and especially with Sergey Pankov. We
thank Heinrich Jaeger and  Thu Tran for many useful discussions
and for providing us with their latest experimental data.  We
thank M.~V.~Fistul for pointing us out on the importance of the
inelastic cotunneling processes.
\end{acknowledgments}

\appendix

\section{Minimization of the probability of the inelastic cotunneling}

   The result for the hopping conductivity in the regime of inelastic cotunneling
   is obtained via  minimization of expression (\ref{result_in1}) with respect to
   the hopping distance $N$ under constraint
   \begin{equation}
   \label{condition}
   N a \Delta \, \tilde \kappa \, / e^2 = b,
   \end{equation}
   following from the ES expression for the density of states,
   with $b $ being a constant.
   Minimization procedure can be easily implemented presenting the tunneling probability $P_{in}$ as
   \begin{equation}
   P_{in} \sim e^A,
   \end{equation}
   with the action $A$ given by
   \begin{eqnarray}
   \label{action}
   A&=&-{{2N} \over \xi_{in}} +\ln \Gamma(N+i\Delta /2\pi T )  \nonumber  \\
   &+& \ln \Gamma (N-i\Delta/2\pi T
   )-\ln\Gamma(2N) -\Delta/2T,
   \end{eqnarray}
   where $\xi_{in}$ is the localization length given by
   (\ref{inelastic_loc_len}).
   Taking into account that according to Eq.(\ref{condition}) $d \Delta/ d N = - \Delta / N$
   minimization of the action $A$ with respect to $N$ in the limit
   of the large $N$ results in the equation
   \begin{equation}
   \label{x}
   \ln 4 + 2/\xi_{in} = \ln(1+x^2)+\pi x \, [ 1+(2/\pi) \arctan x ],
   \end{equation}
   where the parameter $x$ is defined as
   \begin{equation}
   \label{def_x}
   x=\Delta/2 \pi T N.
   \end{equation}
   We see that Eq.(\ref{x}) fixes the ratio $\Delta/T N$ for a
   given localization length. Using  Eq.(\ref{x}) and that $N \gg 1$ the
    action (\ref{action}) can be simplified to
   \begin{equation}
   A=-{\Delta\over { T }} \left( 1+{ 2 \over \pi} \arctan x
   \right),
   \end{equation}
   that with the help of Eqs.(\ref{condition},\ref{def_x}) can be further written in terms the parameter $x$
   as
   \begin{equation}
   A=-\sqrt{{2 \,b\, e^2  }\over {a T \tilde \kappa } }\, \sqrt{x\pi } \; [ 1+ ( 2 / \pi ) \arctan x ].
   \end{equation}

   Within the logarithmic accuracy we can consider the localization
   length as a small parameter.
   In this case Eq.(\ref{x}) gives that $x \approx 1/\pi \xi$
   and from the above equation we finally obtain the resulting
   action
    \begin{equation}
    \label{final_action}
    A \approx -   \sqrt{ {{ 8\,  b \, e^2  }\over { \xi_{in} a\, T \tilde \kappa   }}
    }.
    \end{equation}
    We would like to note that, in principle, one
   should consider an alternative minimization scheme for the tunneling process:
   The tunneling first occurs with $\Delta = 0$ as in the case of
   the elastic process, and then the  electron gets an extra energy due to
   phonon or electron collision not included directly in the
   probability of the inelastic process under consideration. One
   can easily show that in this case one would get $A= - \sqrt{ 8\, b\, e^2\,/T \,a\,\xi_{in} \tilde \kappa   }$
   that coincides with the action (\ref{final_action}) obtained in the
   leading order in the parameter $1/\xi_{in}.$ One can check, however,
   that taking large but finite $1/\xi_{in}$ results in a correction to
   the action (\ref{final_action}) that consist of the reduction of the coefficient
   $\sqrt{8}.$ Thus the process with intrinsic inelasticity that we considered is a
   bit more favorable. This insures that the minimization scheme
   used by us is the right one.


\begin{thebibliography}{99}
\bibitem{footenote1} We consider the system periodic if (i) all the granules are
of the same size and shape and arranged in a regular structure;
(ii) all the tunneling conductances between the nearest neighbors
are equal; and (iii) there is no disorder in the dielectric matrix
that causes local fluctuations in granules' chemical potential.
\bibitem{footenote2} Eq.~(\ref{result_in1}) gives the probaillity of
spin unresolved process, i.e. it includes the sum over spin
components of the final state.
\bibitem{temperature} We use the dimensionless units for energy
of the classical model assuming that the effective temperature
$T=1.$
\end{thebibliography}
\end{document}